\documentclass[iop,numberedappendix,appendixfloats,twocolappendix]{emulateapj}

\usepackage{color}

\shorttitle{Interstellar dust in M51 from AKARI}
\shortauthors{Egusa et al.}

\begin{document}


\title{Interstellar dust properties of M51 from AKARI mid-infrared images}


\author{Fumi Egusa\altaffilmark{1}, Takehiko Wada\altaffilmark{1}, 
Itsuki Sakon\altaffilmark{2}, Takashi Onaka\altaffilmark{2}, Ko Arimatsu\altaffilmark{1,2}, 
Hideo Matsuhara\altaffilmark{1}
}

\email{fegusa@ir.isas.jaxa.jp}




\altaffiltext{1}{Institute of Space and Astronautical Science, 
Japan Aerospace Exploration Agency, Sagamihara, Kanagawa 252-5210, Japan}
\altaffiltext{2}{Graduate School of Science, 
The University of Tokyo, Bunkyo-ku, Tokyo 113-0033, Japan}


\begin{abstract}
 Using mid-infrared (MIR) images of four photometric bands of the Infrared Camera (IRC) 
onboard the AKARI satellite, S7 (7 $\mu$m), S11 (11 $\mu$m), L15 (15 $\mu$m), and L24 (24 $\mu$m), 
we investigate the interstellar dust properties of the nearby pair of galaxies M51 
with respect to its spiral arm structure. 
 The arm and interarm regions being defined based on 
a spatially filtered stellar component model image, 
we measure the arm-to-interarm contrast for each band. 
 The contrast is lowest in the S11 image, which is interpreted as that 
among the four AKARI MIR bands 
the S11 image best correlates with the spatial distribution 
of dust grains including colder components,
while the L24 image with the highest contrast traces warmer dust heated 
by star forming activities.
 The surface brightness ratio between the bands, i.e.\ color, is measured over the disk of 
the main galaxy, M51a, at 300 pc resolution. 
 We find that the distribution of S7/S11 is smooth and well traces the global spiral arm pattern 
while L15/S11 and L24/S11 peak at individual \ion{H}{2} regions.
 This result indicates that the ionization state of PAHs is related to the spiral structure. 
 Comparison with observational data and dust models also supports the importance of 
the variation in the PAH ionization state within the M51a disk.
 However, the mechanism driving this variation 
is not yet clear from currently available data sets.
 Another suggestion from the comparison with the models is that 
the PAH fraction to the total dust mass is higher than previously estimated.
\end{abstract}



\section{Introduction} \label{sec:intro}
 Interstellar dust is abundant and widely distributed 
especially within a disk of late type galaxies.
 Since the properties of dust reflect conditions of its environments, 
emission from dust is an important probe for interstellar medium (ISM).
 Dust around star forming regions absorbs intense star light and radiates 
the absorbed energy in the infrared (IR) regime.
 Given this close connection between the IR emission and the UV-optical extinction, 
the 24$\mu$m band flux has recently been used to correct for the extinction effect 
on the star formation rate derived from the optical H$\alpha$ emission \citep{Cal07}.
 At shorter wavelengths, many spectral band features are observed and 
polycyclic aromatic hydrocarbons (PAHs) have been regarded to be their major carriers
\citep[e.g.,][]{LP84,Sell84,Alla89,Dra03,Tie08}.
 These features at different wavelengths are thought to represent PAHs in different conditions \citep{DL07}.
 For example, the 7.7$\mu$m feature, which is bright in star forming sites, mostly originates from 
ionized PAHs, while neutral and ionized PAHs equally contribute to the 11.3$\mu$m feature. 
 Features at shorter wavelengths are generally emitted from smaller grains, and vice versa.
 Ratios of these PAH band features are thus used to investigate its environment and property 
such as temperature and ionization state \citep[e.g.,][]{Gal08, Mori12}.

 M51 is a pair of the grand-design spiral galaxy, M51a (NGC 5194) 
and the companion galaxy, M51b (NGC 5195).
 With its proximity and prominent two-armed spiral structure, 
it has been a popular target to study ISM 
and its relationship to the spiral arm pattern.
 From optical and UV images, properties of \ion{H}{2} regions and star clusters for the entire disk of M51a 
have been investigated, 
and their ages in the arms are found to be younger than those in interarm area \citep{LeeJ11, SanG11}.
 Their natal clouds are traced by radio CO lines and the largest clouds, which are thought to be a 
progenitor of massive star clusters, are only found in the arms \citep{Koda09}.
 These results indicate that star formation dominantly occurs in spiral arms.
 \citet{Egu11} investigated the evolutionary sequence of molecular gas {\it within} an arm 
using very high angular resolution CO data and found that 
cloud cores, which eventually form stars, are formed through the passage of spiral structure.
 On the other hand, studies on 
warm and hot interstellar dust components 
with mid-IR (MIR) observations
are rather limited.
 For selected locations, \citet{Sauv96} analyzed 7$\mu$m (LW2) and 15$\mu$m (LW3) images of 
M51 of the ISO/ISOCAM and found that LW2/LW3 becomes small in the spiral arms, 
which they attribute to a temperature variation.
 From the MIR spectra toward a central $\sim 3'\times 3'$ area of M51a, 
\citet{Gal08} derived the PAH band ratio of 7.7$\mu$m/11.3$\mu$m to be higher in the spiral arms.
 These results indicate warm and ionized dust in spiral arms but the variation of dust properties 
across the entire disk has not yet been investigated.

\begin{figure}
\plotone{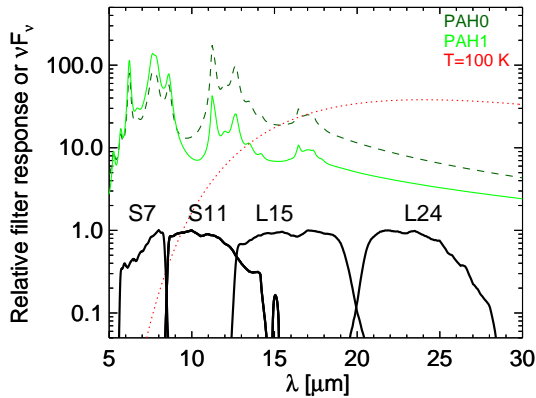}
\caption{Relative MIR filter response curves normalized to unity at peak (thick solid black) 
together with spectra for neutral PAH (PAH0; dashed dark green), ionized PAH (PAH1; solid light green), 
and modified black body with $T=100$ K (dotted red).}
\label{fig:rsr}
\end{figure}

 The Infrared Camera \citep[IRC;][]{IRC} on board the AKARI satellite \citep{AKARI} has 
an advantage in the wide field of view of $\sim 10'$, comparable to the size of the whole M51 system.
 In addition, three channels, NIR, MIR-S, and MIR-L, and three bands in each channel
cover the wavelength range of 2--25 $\mu$m {\it continuously}.
 In Figure \ref{fig:rsr}, relative response curves of MIR bands used in this study
(S7 and S11 for MIR-S and L15 and L24 for MIR-L, 
while the numbers after S or L denote its representative wavelength in $\mu$m) 
are plotted together with template spectra for neutral PAH, ionized PAH, and 
dust grain with $T=100$ K.
 PAH spectra are from \citet{DL07} provided by DustEM \citep{Com11}, 
and the modified black body spectrum is used to represent the $T=100$K 
dust component.
 These spectra are also used in this study and fully described in \S \ref{sec:cc}.
 As illustrated in this figure, AKARI MIR 
bands efficiently trace the interstellar dust in different conditions, 
helping us to reveal dust properties comprehensively.
 The angular resolution, 
i.e.\ the Full Width at Half Maximum (FWHM) of a Point Spread Function (PSF), 
is $4.0''$--$6.8''$, depending 
on the observing wavelength, and is high enough to investigate structures within the galactic disk.
 With these characteristics, AKARI images provide valuable information 
for understanding of the interstellar dust condition and its relationship to the 
environments such as star forming activities and galactic structures.

 In this study, we use AKARI MIR images to understand dust properties in M51 
and to investigate their spatial variation to what extent 
the grand-design spiral structure has an effect.
 We present that the 11 $\mu$m image is of great importance to explore 
PAH conditions such as ionization state and its fraction to total dust mass.
 Adopted parameters of M51a are listed in Table \ref{tb:m51}.
 At the distance of 8.4 Mpc, $1''$ corresponds to 41 pc.

\begin{deluxetable}{ccc}
\tablecaption{Parameters of M51a\label{tb:m51}}
\tablewidth{0pt}
\tablehead{ \colhead{Parameter} & \colhead{Value} & \colhead{Reference}}
\startdata
RA (J2000) & 13:29:52.711 & NED\\
dec (J2000) & +47:11:42.62 & NED\\
PA (deg) & $169.0\pm 4.2$ & \citet{Dai06} \\
inc (deg) & $24\pm 3$ & \citet{Shet07} \\
distance (Mpc) & 8.4 & \citet{Fel97}
\enddata
\tablecomments{NED centers, originally from radio continuum observations \citep{TH94}, 
are $< 1''$ from dynamical centers derived by \citet{Shet07}.}
\end{deluxetable}

\section{Data Reduction} \label{sec:red}

\subsection{IRC data sets}
 By design, NIR and MIR-S share the field of view while 
MIR-L is offset by $\sim 25'$.
 Each covers about $10' \times 10'$, which is almost 
the same as the M51 system in size.
 Properties of the data sets used in this study are summarized in Table \ref{tb:obs}.
 They were taken as a part of the mission program, 
``ISM in our Galaxy and Nearby Galaxies'' \citep[ISMGN;][]{Kane09}, 
and the astronomical observation template (AOT) 02 was used.
 In this AOT, two bands in each channel were used and 
one observing cycle is one short and one long exposures for NIR, 
while one short and three long exposures for MIR.
 A slight position change (dithering) was performed every cycle 
to minimize the effect of unusable pixels due to hot pixels and cosmic rays.
 
 The data sets were retrieved through Data ARchives and Transmission System
(DARTS)\footnote{http://darts.isas.jaxa.jp/astro/akari/akarilog/top.do}, and 
raw images were processed with the latest IRC imaging toolkit, 
a package of the Image Reduction and Analysis Facility (IRAF)\footnote{IRAF is distributed 
by the National Optical Astronomy Observatories, 
which are operated by the Association of Universities for Research in Astronomy, Inc., 
under cooperative agreement with the National Science Foundation.} scripts available via the 
AKARI observers page\footnote{http://www.ir.isas.jaxa.jp/ASTRO-F/Observation/}.

 In the following part of this section, we briefly describe the reduction procedures 
that are not included in the standard pipeline but are necessary for the M51 data.
 Detailed descriptions for some of these procedures are presented in a separate paper 
(Egusa et al.\ 2013, in prep.).

\begin{deluxetable}{cccc}
\tablecaption{Properties of M51 AKARI/IRC data sets\label{tb:obs}}
\tablewidth{0pt}
\tablehead{ \colhead{ID} & \colhead{Obs date} & \colhead{Filters} 
& \colhead{short, long}}
\startdata
1400387.1 & 2006/12/16 & N3, N4 & 4, 4\\
 & & S7, S11 & 4, 12\\
1402202.1 & 2007/06/16 & L15 & 3, 9\\
 & & L24 & 4, 12
\enddata
\end{deluxetable}

\subsection{Neighbor dark frames}
 Since the observations on orbit started, 
the number of hot and warm pixels has increased with time especially in the MIR-L channel.
 The super dark frames were created from the data sets taken 
at the very early stage of the satellite operation and thus are not 
best suited to data sets taken at later stages.
 A few dark frames for each channel and exposure were taken at the beginning and the end 
of each pointed observation and the self dark frames are generally 
created from so-called pre-dark frames taken prior to the object frames.
 They are however less reliable due to the small number of frames combined.
 
 We thus created a new set of dark frames called neighbor dark.
 The concept is combining dark frames taken in a certain period of time 
in order to increase the number of data combined (than self dark) 
and to better calibrate temporal variations (than super dark).

\subsection{Earth shine light removal}
 The background of MIR-L frames of M51 was found to be 
contaminated by the earth shine light (EL).
 Since the EL pattern is extended, 
it is difficult to be separated from the object, M51.
 In addition, its amplitude varied significantly during the observations 
as the separation angle between the telescope and the earth changed.
 We have thus created an EL template for each MIR-L band 
from data sets taken under similar conditions (i.e.\ date and coordinates), 
and successfully removed the EL component from the MIR-L frames. 
 Scripts for creating and removing the EL template are also included in the IRC toolkit.

 In the case of M51, the maximum of EL contamination was 
about 2 and 30 MJy/sr for L15 and L24, respectively.
 The former is comparable to faint diffuse emission in interarm regions, 
while the latter reaches to the brightness of \ion{H}{2} regions in spiral arms.

\subsection{Stacking}
 For the sake of PSF correction, 
the pixel scale of MIR images was made half, 
i.e.\ one pixel was divided into $2\times 2$ pixels, before stacking.

 For MIR-L images, only few stars were available within the field of view 
to calculate shifts between the frames.
 The reduction toolkit picked up objects in M51, mainly bright spots in spiral arms, 
instead.
 In order to check the validity of shift values derived, we used the IRAF task 
{\tt xregister}, which calculates a cross correlation between two images 
and determines relative shifts in x- and y-direction.
 An area for calculating the cross correlation was carefully selected not to include masked pixels.
 A script to perform this procedure is also included in the IRC toolkit 
as an alternative when the normal coadd process fails, and especially useful for 
diffuse objects observed in the MIR-L channel.

 In the case of MIR images of M51, we confirmed that the shift values derived by 
the normal coadd and {\tt xregister} are consistent within 0.5 pixel.

 Only the frames with long exposure were stacked.

\subsection{WCS registration}
 For NIR images, the World Coordinate System (WCS) information 
was automatically set by the IRC toolkit, 
which uses the 2MASS point source catalog.
 For MIR images, on the other hand, the number of stars in the field of view was not enough 
to derive the accurate coordinates for the toolkit.
 We thus picked up stars in the stacked image manually and obtained their coordinates 
in the USNO catalog with the help of the Aladin software.
 For MIR-S images, about five stars were found to be useful, while 
only two stars were found in the field of view of MIR-L.
 As a consequence, we had to add the nucleus of M51a.
 Based on the position of these stars in the image plane and their coordinates, 
the WCS header parameters were derived and updated with the use of 
the IRAF task {\tt ccmap}.
 In order to assess the validity of this procedure, we derived the coordinates 
of the N4 image and confirmed that they are consistent with the toolkit-derived 
coordinates.
 The rms of the coordinate fitting was $\sim 1''$ for NIR and MIR-S images 
and $\sim 2''$ for MIR-L images.
 With all the six images whose WCS was set, we confirmed their coordinates 
are consistent between the bands and also with other archival images.

\subsection{PSF correction}
 In order to discuss brightness ratios between different bands, 
we applied the PSF correction to the four MIR band images. 
 The PSF template for each MIR band is provided by \citet{Ari11}.
 The corrected PSF is 2D gaussian with FWHM of $7.4''$, 
which is slightly larger than the original L24 PSF 
and corresponds to $\sim 300$ pc at the distance of M51.

\subsection{A ghost in MIR-S}
 A bright compact object close to M51b in the south-west direction, 
which is apparent only at S7 and S11, 
was found to be an artifact called ghost of this companion galaxy, 
arising from internal reflection in the beam splitter.
 The peak flux of the ghost relative to M51b is larger at S11.
 \citet{Ari11} examined major artifact features including the ghost in MIR images and 
provided their template patterns for each band. 

 We tried to remove this ghost component from the stacked S11 image before 
the PSF correction in two different ways. 
 We first made a model ghost pattern from the flux distribution of 
the companion galaxy by convolving it with the ghost template.
 Since the model did not fit the observed pattern perfectly, 
we then tried to remove it by employing the CLEAN algorithm \citep{Hog74}, which has been 
used for deconvolution of radio interferometer images.
 A CLEANed image however showed a residual pattern, which cannot be neglected.
 
 We thus concluded that the ghost template is not applicable to M51 S11 image.
 The reason is plausibly a combination of the following facts.
 First, the template is made for a bright source at the center of the field of view. 
 A spatial pattern for an object around the edge, which is the case for M51b, can be different.
 In fact, the peak position of the observed ghost is slightly off from 
the expected position.
 Second, the ghost amplitude is known to depend on the source color \citep{Ari11}.
 Third, the count of the source, M51b, is very high and close to the saturation limit.
 For such a bright source, the uncertainty in the linearity correction would be rather large.
 The latter two facts can explain the failure in modeling the ghost, but cannot in CLEANing.
 
 We therefore decided to mask a circular area with radius of $15''$ 
around the observed ghost in S11.
 The same area was also masked in S7.
 The masking was performed after the PSF correction.

\subsection{Regridding \& flux calibration} 
 The flux conversion factors provided by \citet{Tana08}
were applied to our data sets.
 The $1\sigma$ uncertainties of the conversion factors 
are estimated to be $\sim 2$\% for S7 and S11, 
$\sim 3$\% for L15, and $\sim 5$\% for L24.
 Color correction was not applied.
 We note here that a sky background was subtracted before flat fielding.
 
\begin{figure*}
\plotone{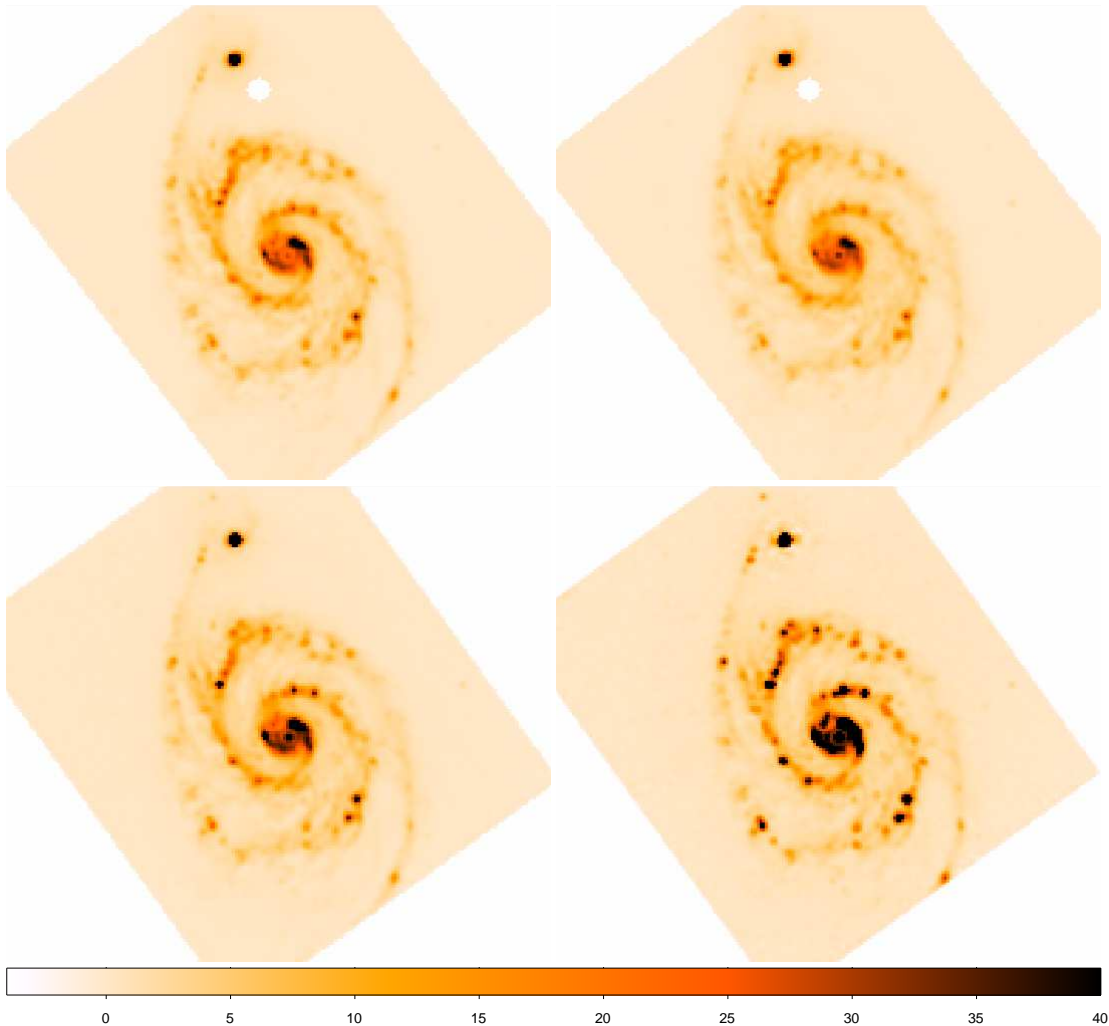}
\caption{PSF corrected, aligned, and calibrated images for 4 MIR bands, 
S7 (top left), S11 (top right), L15 (bottom left), L24 (bottom right). 
The FWHM of the corrected PSF is $7.4''$ and the pixel size is $3.7''$.
The unit for color scale is MJy/sr.
The ghost has been masked in S7 and S11.
The stellar contribution is not subtracted.}
\label{fig:MIR4}
\end{figure*}

 All images were aligned North up and East left, and regridded to $3.7''/{\rm pix}$ (Figure \ref{fig:MIR4}).

 Surface brightness distribution at S7, L15, and L24 was compared with 
that of Spitzer IRAC4, ISO LW3, and Spitzer MIPS24, respectively.
 The PSF matching was performed if necessary.
 We found that the corresponding two data sets are linear with a slope 
around unity.

\section{Analysis}
\subsection{Stellar component distribution}\label{sec:submodel}
 While dust is the dominant source of MIR emission in galaxies, 
stellar contribution is sometimes not negligible at shorter wavelengths 
and highly depends on the total stellar mass, the star formation history, and so on.
 Fitting stellar SEDs generated by the PEGASE.2 code \citep{FIoc97} 
to optical and NIR images of M51, \citet{MenC12} derived a set of best-fit parameters 
of the stellar population for each pixel. 
 Based on this model of stellar distribution, 
the contribution from stars to the AKARI MIR bands was calculated 
and subtracted from individual band images so that they represent dust-only properties.
 The stellar fraction to the total MIR flux was typically a few percent and 
higher values ($\sim 20$\%) were found in a part of spiral arms at shorter wavelengths.
  
 In addition to stars in M51, foreground MW stars were visible in some AKARI images.
 In order to identify such foreground stars, we performed the aperture photometry on the S7 image
of sources listed in 2MASS Point Source Catalog within the FoV 
and compared the S7 flux with $K$ magnitude.
 We manually set a border between foreground stars and 
extragalactic sources in the S7 -- $K$ plot 
and defined $\sim$ 20 sources as plausible foreground stars.
 Pixels within $7.4''$ from these stars were masked out.

 All the results and discussions presented in \S \ref{sec:res} and \S \ref{sec:dis} 
are based on the dust only images 
created from stellar subtraction and masking described in this subsection.

\subsection{Arm/interarm definition}\label{sec:armdef}

 In this study, we basically followed the procedures of \citet{Dum11}, 
who applied the wavelet analysis to the 20 cm radio continuum image of M51 
to separate arm and interarm regions.
 Since stars dominate the baryonic mass and thus the gravitational potential of the galactic disk, 
we used the model stellar distribution 
by \citet{MenC12} mentioned above to define arm and interarm regions.

 We first deprojected the image to obtain a face-on view of M51 with 
the disk orientation parameters listed in Table \ref{tb:m51}.
 An azimuthal average was calculated at each radius 
and subtracted to remove axisymmetric components and thus 
to enhance the asymmetric spiral arm structure.
 The central region with $r \leq 30''$ from M51a, 
the area within $80''$ from M51b, and the foreground stars 
defined in the previous subsection were masked out.

 Following \citet{Dum11}, we then applied the wavelet analysis with Pet Hat function 
in order to extract a specific spatial scale.
 The result was consistent with \citet{Dum11}, the scale of $83''$ (or 3.4 kpc) being suitable to delineate 
the two-armed spiral structure of M51a disk.
 Our definition appears to be smoother than theirs 
since we used the stellar distribution, 
while they used the 20 cm image, tracing the synchrotron emission from supernova remnants 
and diffuse ISM and thus being more sensitive to smaller structures.

\begin{figure}
\includegraphics[width=\linewidth,trim=60 40 40 60,clip]{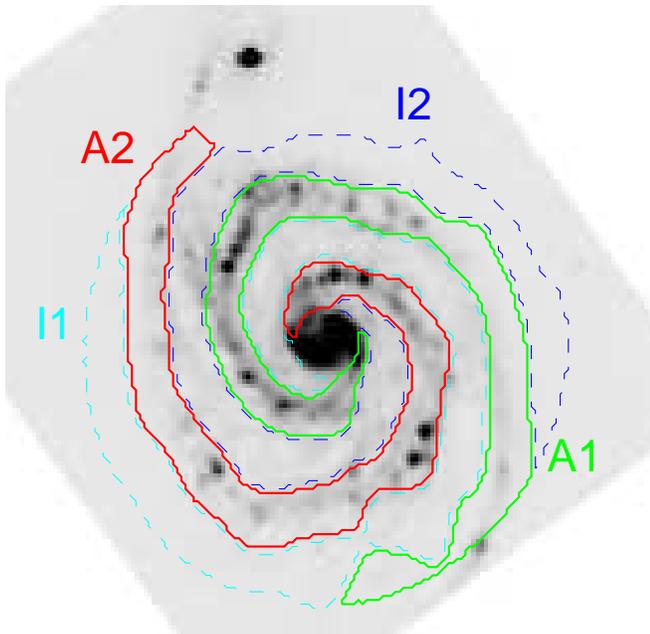}
\caption{Arm and interarm definition of M51a.  
Areas enclosed by solid green and red lines 
are defined as arms (labeled as A1 and A2), 
while those with cyan and blue dashed lines are interarm regions (labeled as I1 and I2).
The background is the L15 image.}
\label{fig:armdef}
\end{figure}

 The minimum and maximum radius of the disk was set to be $30''$ and $240''$, respectively. 
 The area within $80''$ from the companion was also excluded as mentioned above.
 We separated the disk into arm and interarm regions according to the sign of 
the wavelet component map of $83''$ scale.
 Plus and minus correspond to arm and interarm, respectively.
 The area not connected to the central region was also excluded.
 Following \citet{Egu09}, Arm 1 was defined as extending in the opposite direction to the companion, 
while Arm 2 was defined to be connected to the companion.
 This arm and interarm definition is shown in Figure \ref{fig:armdef}.
 Pixels outside the defined area are excluded in the following.

\section{Results} \label{sec:res}
\subsection{Flux distribution and arm/interarm contrast}
\begin{figure}
\includegraphics[width=\linewidth]{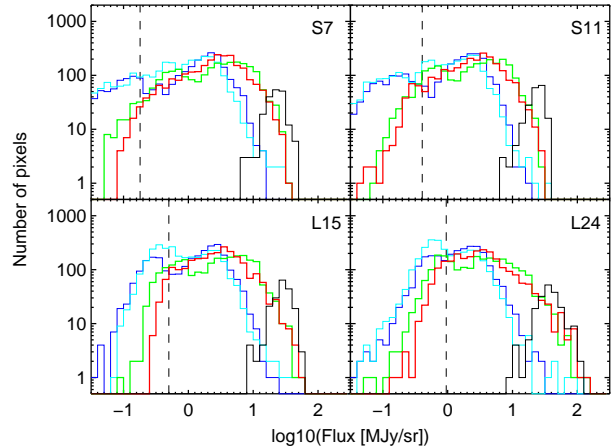}
\caption{Surface brightness histogram of each region for each band. 
Colors correspond to the arm definition shown in Figure \ref{fig:armdef}, 
and black is for the central region.
The vertical dashed line indicates $5\sigma$ of each band.}
\label{fig:fcomp}
\end{figure}

 The four MIR images shown in Figure \ref{fig:MIR4} are overall similar, 
in the sense that they are bright in the center of the two galaxies and in the spiral arms of M51a.
 In Figure \ref{fig:fcomp}, 
the surface brightness histogram for each region defined in \S \ref{sec:armdef}
and each AKARI MIR band is presented. 
 The histogram for the central region is distinct with 
the largest value of the peak and the narrowest width of the profile.
 The histograms for the arm and interarm regions, on the other hand, show 
broader profiles with peaks around 1.0--10 MJy/sr.
 These broader profiles are likely due to 
the definition of the arms, which includes
bright \ion{H}{2} regions and fainter area, 
and also to interarm filamentary structures, which are called feathers or spurs 
and are often bright as arm structures
(See \S\ref{sec:S7/S11} for discussion in more detail).

 We have calculated the brightness ratio of arm to interarm regions, 
i.e.\ the arm/interarm contrast, 
at each $10''$ width annulus.
 It varies from 1 to 5, with higher values found 
where bright star forming regions belong to the arms.
 The mean values of the arm/interarm contrast 
are 2.9 for S7, 2.5 for S11, 2.7 for L15, and 3.1 for L24.
 The highest contrast for L24 indicates that it best traces the warm dust 
around star forming regions in the arms.
 Although S7 is expected to trace star forming activities, its arm/interarm contrast is 
lower than that of L24 since S7 is also bright in 
feathers.
 On the other hand, S11 exhibits the lowest contrast, 
implying that flux around 11 $\mu$m is less sensitive to the environments.
 Consistently, \citet{Ona07} found a constant IRAS 12$\mu$m/FIR ratio 
over a wide range of temperature. 
 For elliptical galaxies where properties of dust grains and their heating sources 
can be different from those in spiral galaxies,
\citet{Kane07} found a good correlation between 
the equivalent width of PAH $11\mu$m feature and FIR/MIR ratio.
 These results suggest that the $11\mu$m brightness 
correlates well with the total amount of dust including colder components, which dominate the FIR flux.
 L15 seems to be a combination of S11 and L24, with diffuse dust distribution similar to S11 
and a few bright spots corresponding to the nuclei and brightest \ion{H}{2} regions.

\subsection{Color distribution}\label{sec:colors}
\begin{figure*}
\includegraphics[scale=0.65,trim=60 10 60 30,clip]{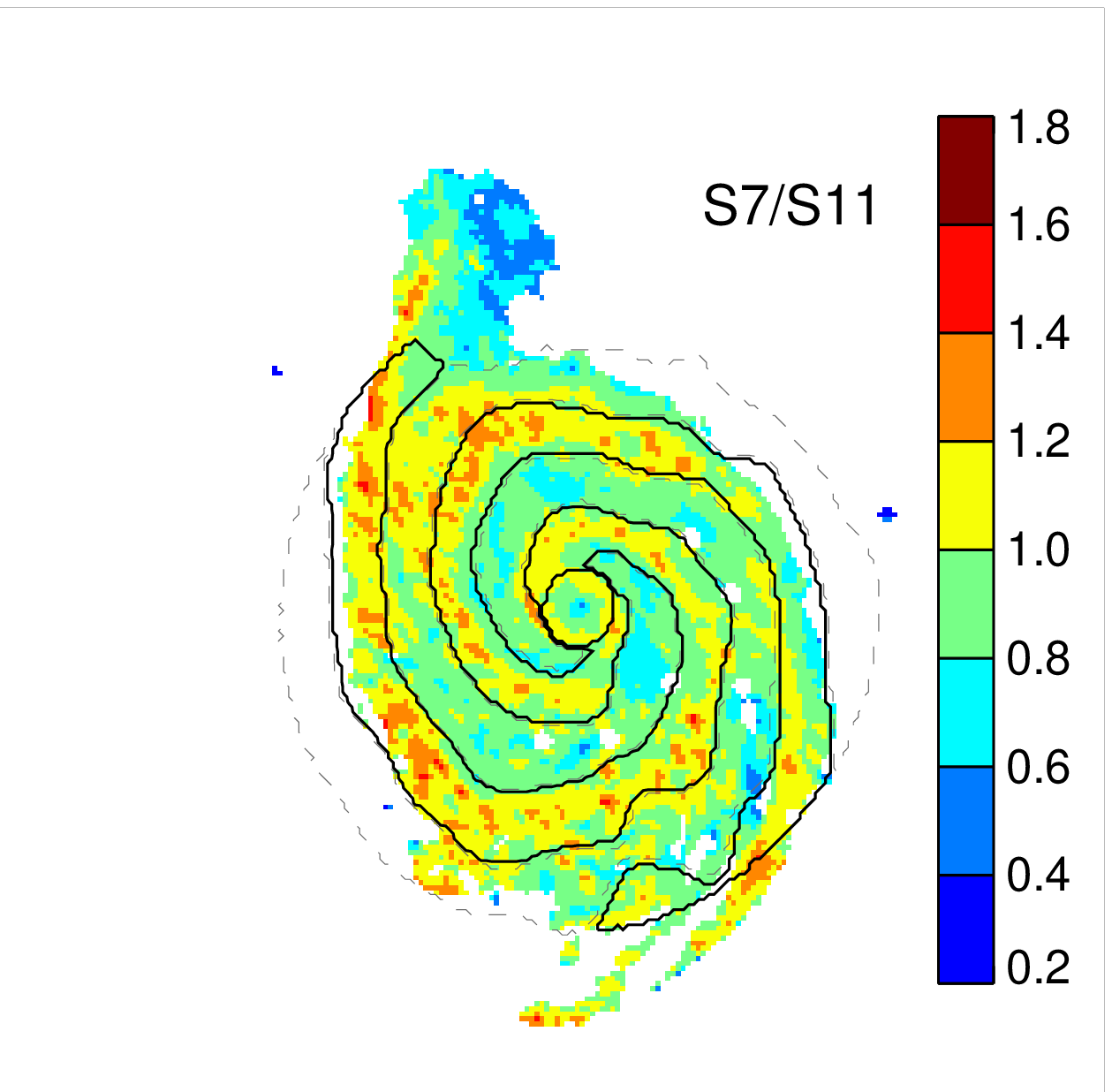}
\includegraphics[scale=0.65,trim=60 10 60 30,clip]{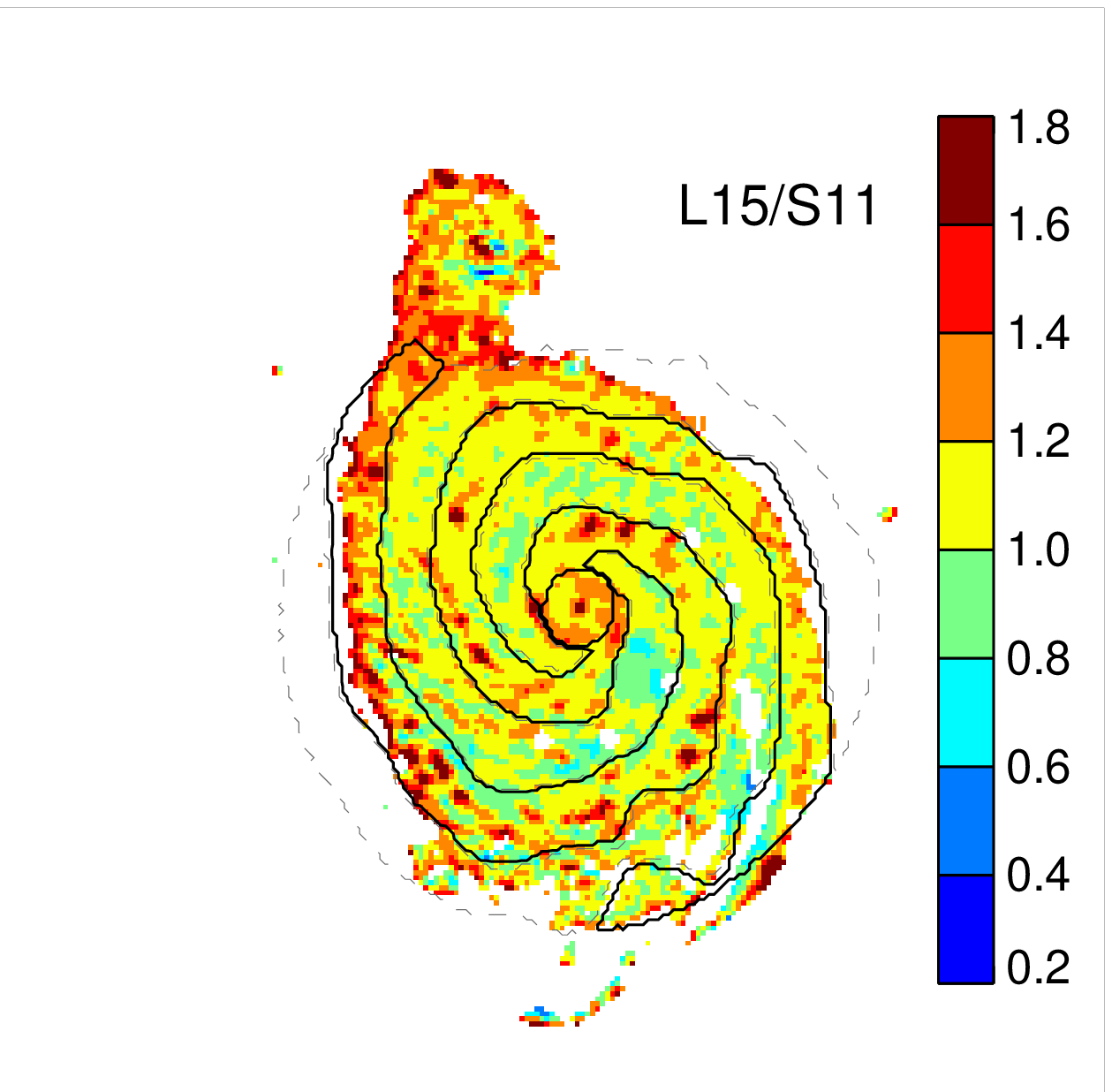}
\includegraphics[scale=0.65,trim=60 10 00 30,clip]{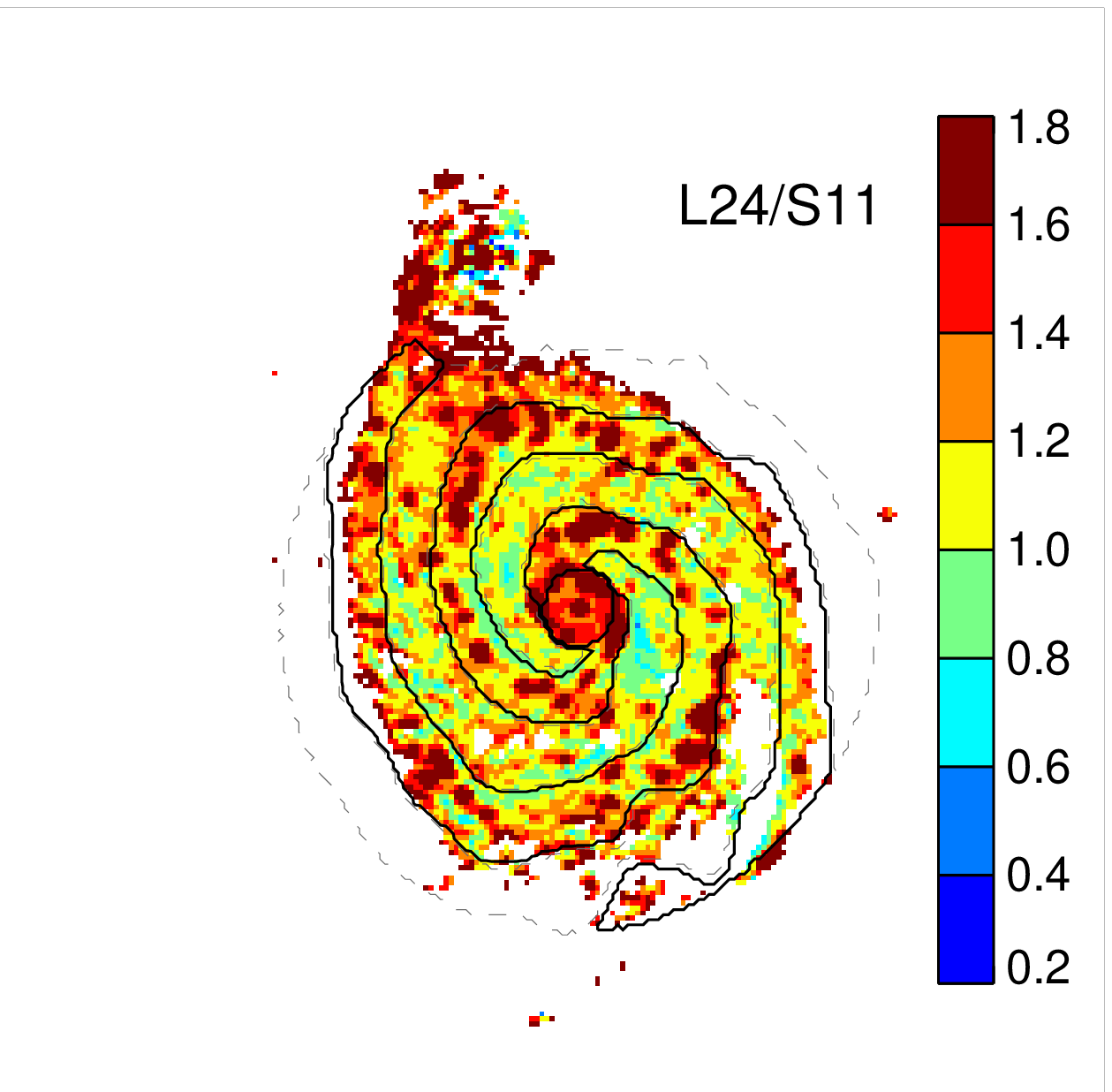}
\caption{Map of colors: (left) S7/S11, (middle) L15/S11, (right) L24/S11. 
The arm and interarm regions defined in \S \ref{sec:armdef} are indicated by thick solid black 
and thin dashed gray lines, respectively.}
\label{fig:color_wa}
\end{figure*}

 A surface brightness ratio map between two bands (i.e.\ color map) of M51 
was created from two band images using all the pixels 
brighter than $5\sigma$ in both images.
 Colors based on S11, i.e.\ S7/S11, L15/S11, and L24/S11, are presented in Figure \ref{fig:color_wa} 
with the defined arm and interarm regions.
 S11 was selected to be the denominator because it shows the lowest arm/interarm contrast 
and thus is thought to best represent the total dust distribution 
among the four AKARI MIR bands.
 Note that S/N of color is lower at larger radii where it is close to the boundary.
 The high L15/S11 values at the east end may be artificial due to residuals of the EL subtraction.
 In addition, L24/S11 around M51b is erroneous due to residuals from the PSF correction.

\begin{figure*}
\includegraphics[width=.33\linewidth]{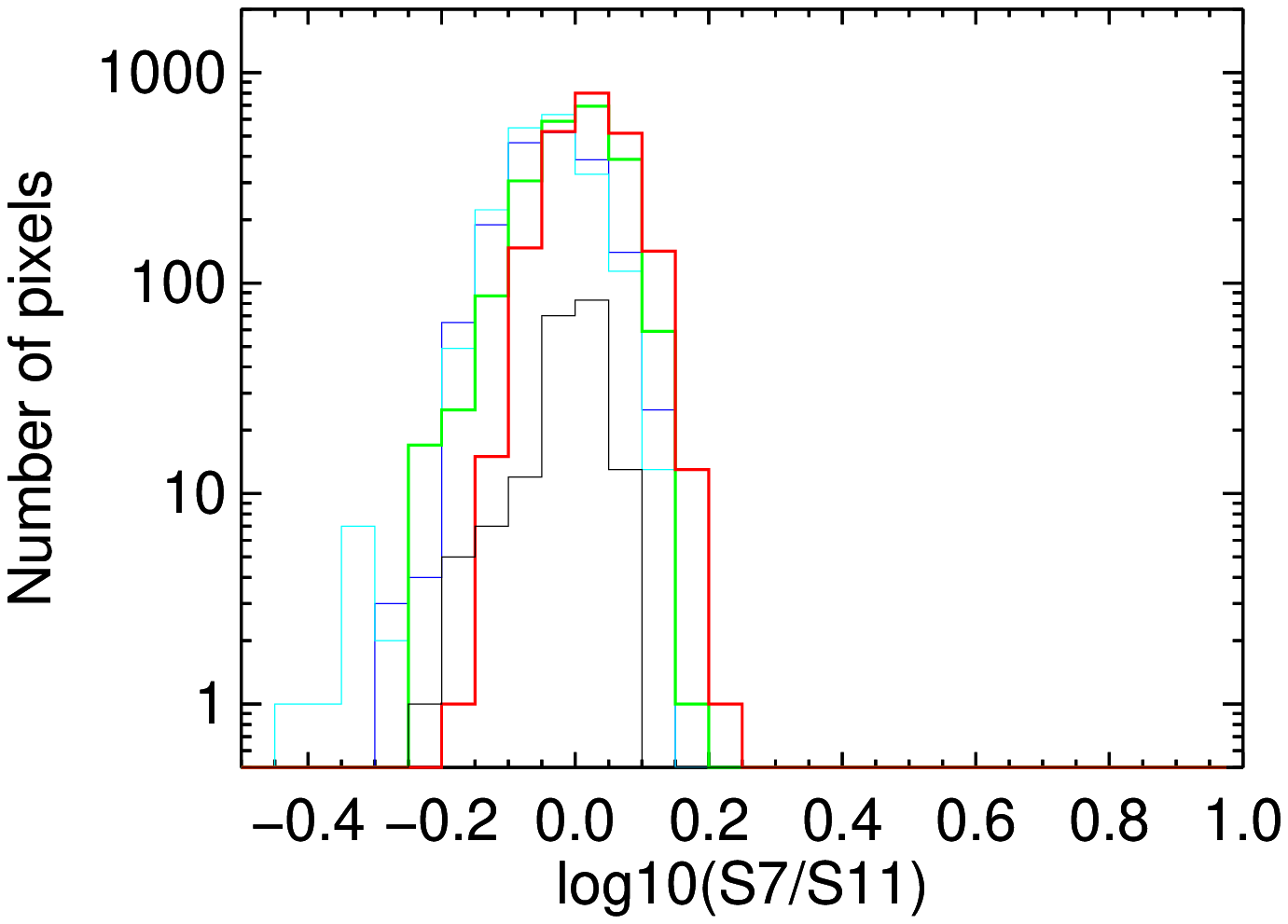}
\includegraphics[width=.33\linewidth]{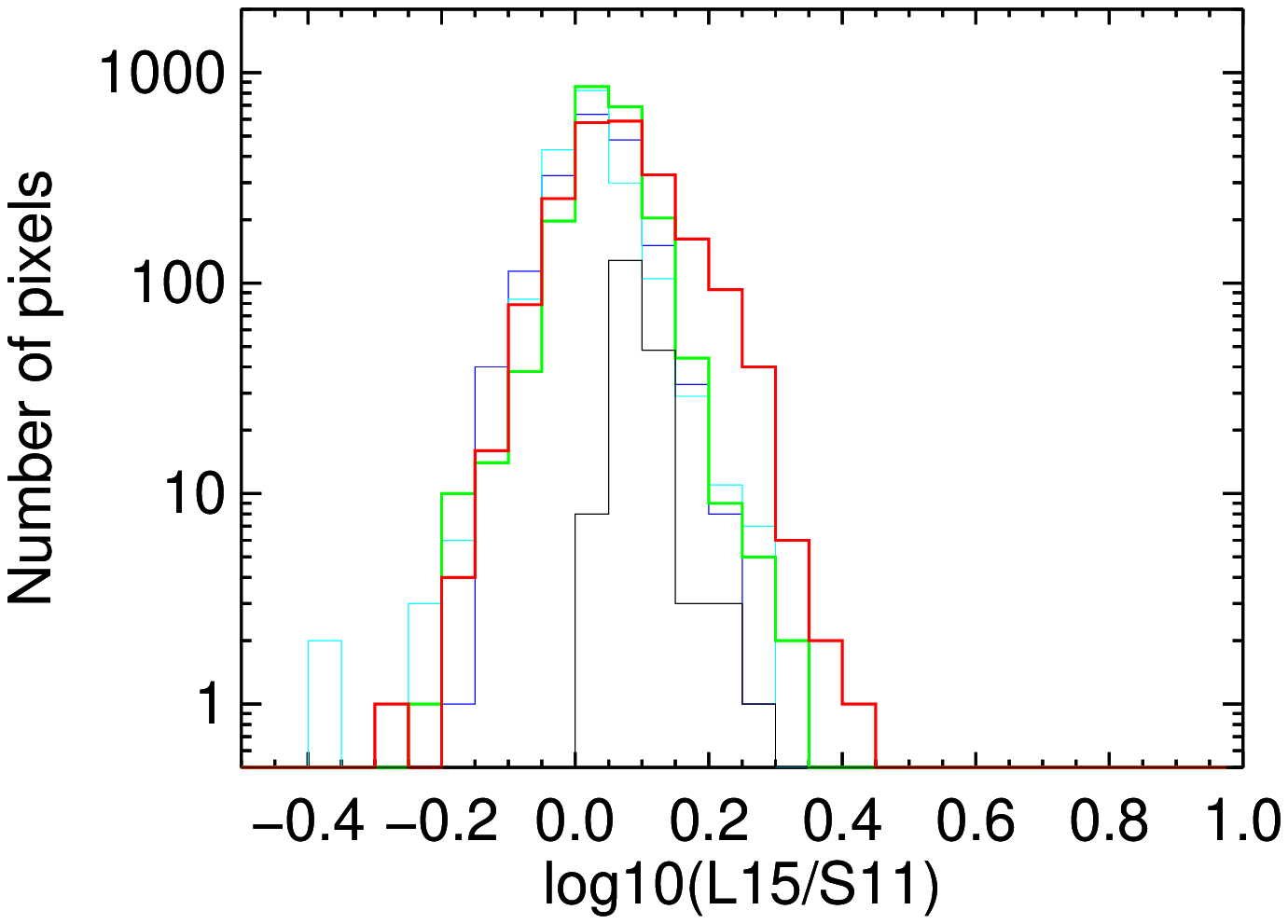}
\includegraphics[width=.33\linewidth]{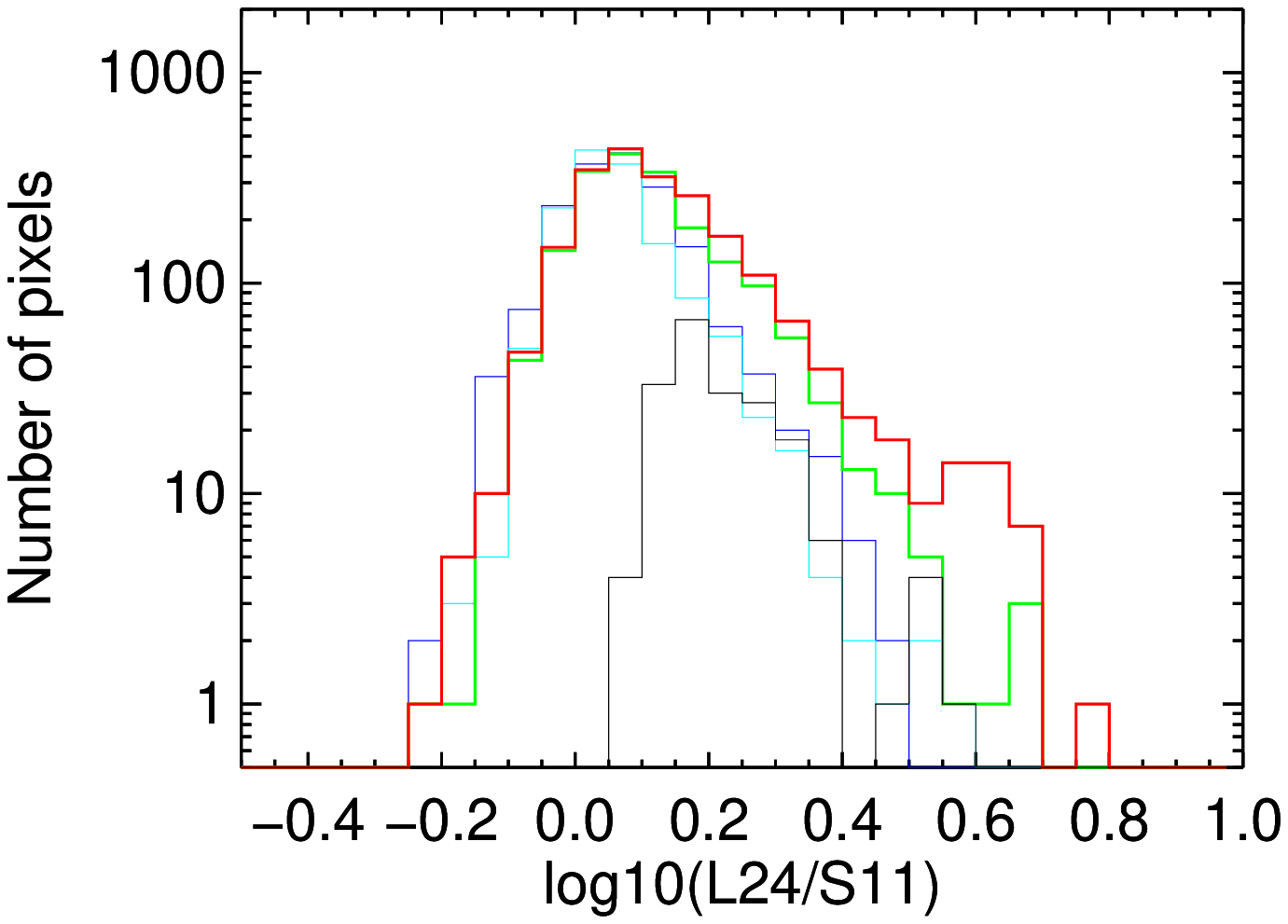}
\caption{
Histograms of colors based on S11 
for the defined regions 
with the same color coding as Figure \ref{fig:fcomp}.
}
\label{fig:chist}
\end{figure*}

 In Figure \ref{fig:chist}, histograms of colors are presented for the defined regions.
 Difference between the regions are less distinct 
compared to the histograms of surface brightness (Figure \ref{fig:fcomp}).

\begin{figure*}
\includegraphics[width=.33\linewidth]{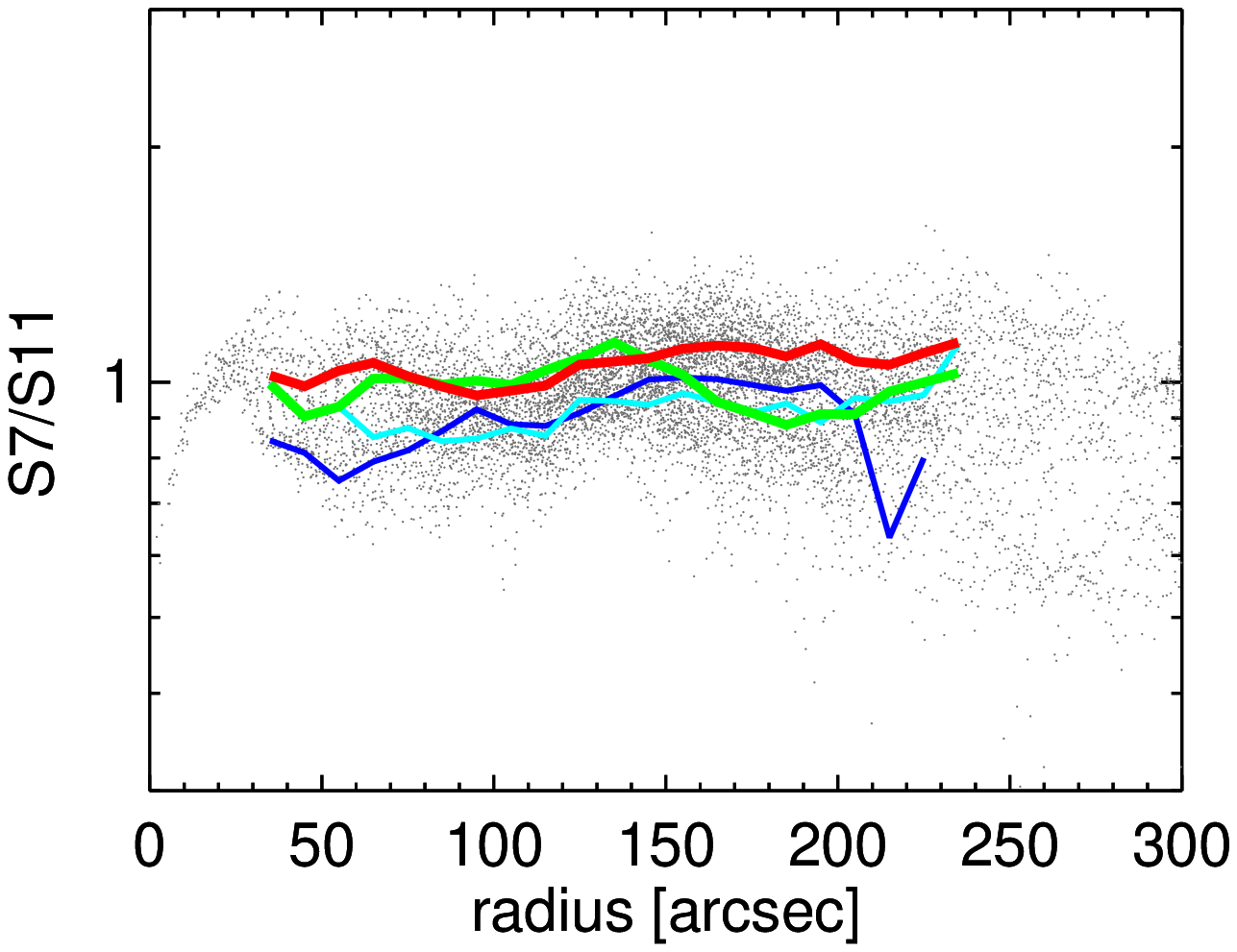}
\includegraphics[width=.33\linewidth]{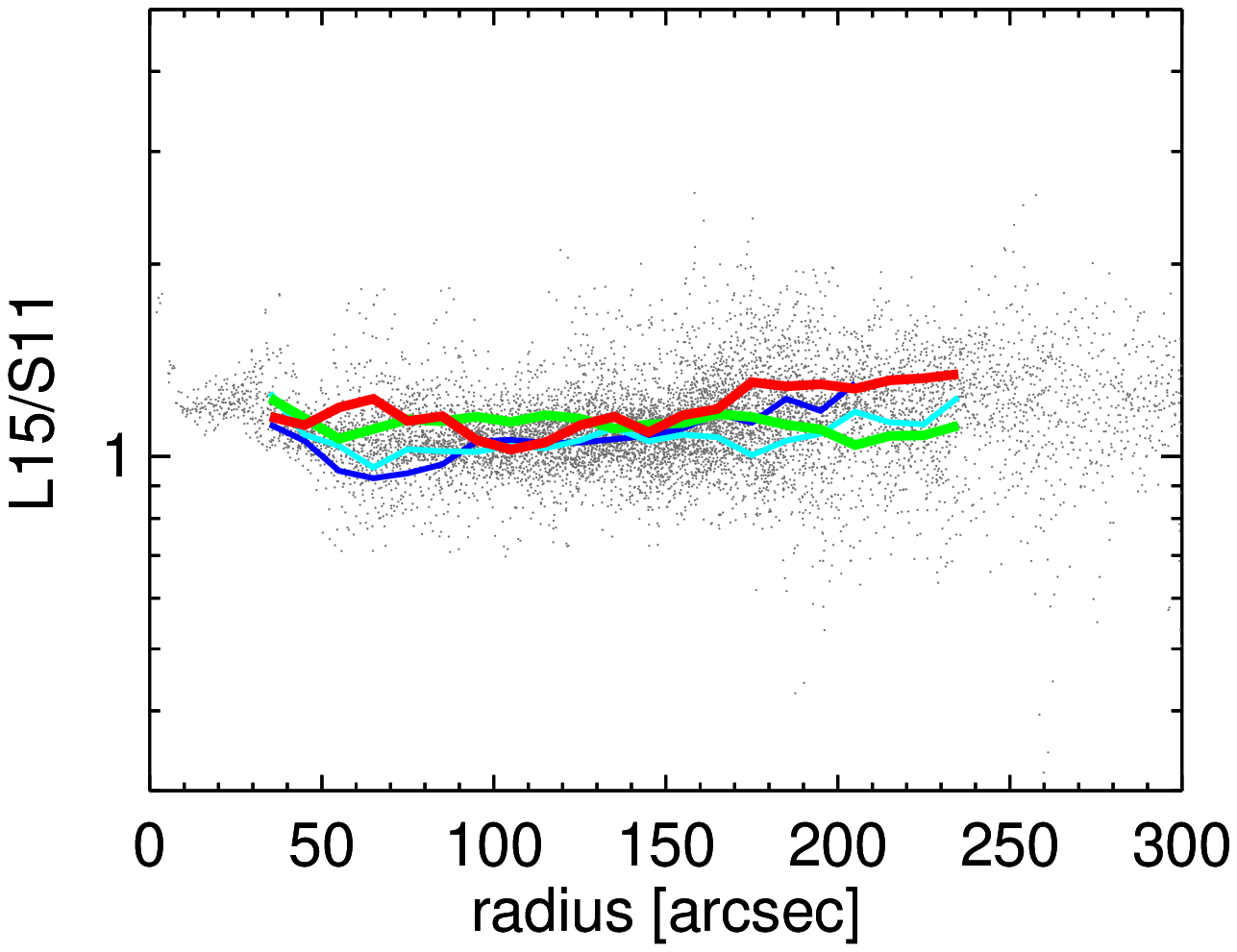}
\includegraphics[width=.33\linewidth]{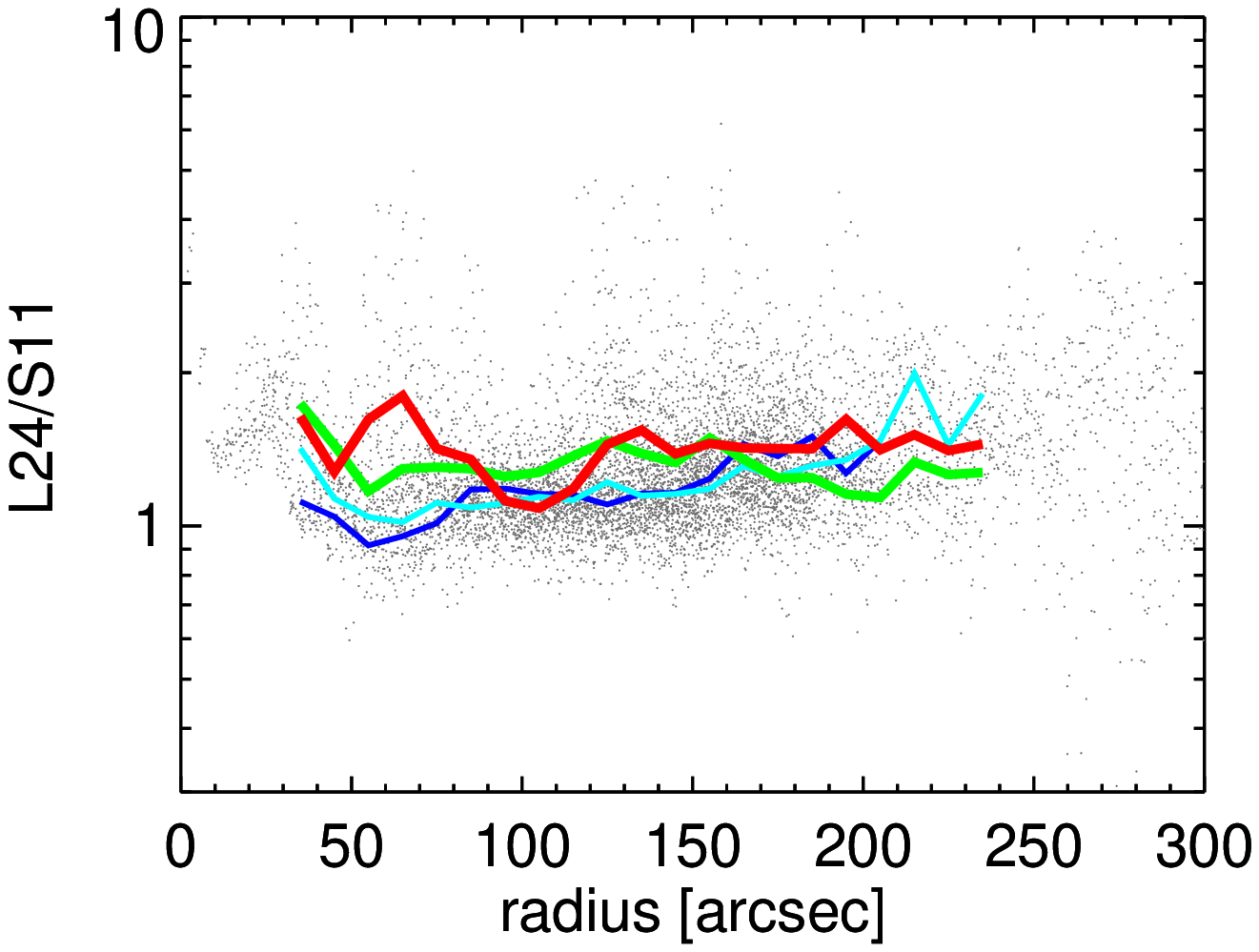}
\caption{Radial profiles of colors based on S11. 
Azimuthal average at every $10''$ for each region is overplotted 
with the same color coding as Figure \ref{fig:armdef}.}
\label{fig:color_rad}
\end{figure*}

 In Figure \ref{fig:color_rad}, these colors are plotted against radius.
 Averages of colors are calculated within an annulus of $10''$ radial width 
and overplotted with lines color-coded in the same way as Figure \ref{fig:armdef}. 
 We find that they are mostly flat, 
indicating that the large dispersion seen in the histogram (Figure \ref{fig:chist}) 
is not due to a radial gradient of colors.
 Using Spitzer IRAC and MIPS photometric data sets, 
\citet{Muno09} constructed spectral energy distributions (SEDs) at representative radii and 
fitted a dust model from \citet{DL07} to the SED at each radius.
 For M51a, a radial variation of the PAH fraction and average radiation field 
is estimated to be small within the disk, while the fraction of the total IR (TIR) luminosity 
contributed by bright star forming regions decreases with radius.
 The small radial variation of the former two parameters is consistent with our flat color profiles.
 On the other hand, L24/S11 is expected to decrease with radius, 
since the latter parameter is related to the $24\mu$m/TIR ratio and 
the $11\mu$m brightness is regarded to correlate with TIR.
 This apparent inconsistency indicates that S11 is not a perfect substitution for TIR.
 The average and rms of each color have been calculated 
for each region and are listed in Table \ref{tb:colors}.

\begin{deluxetable}{cccc}
\tablecaption{Colors based on S11 for each region\label{tb:colors}}
\tablewidth{0pt}
\tablehead{ \colhead{Region} & \colhead{S7/S11} & \colhead{L15/S11} & \colhead{L24/S11}}
\startdata
Center & $0.9 \pm 0.1$ & $1.2 \pm 0.1$ & $1.7 \pm 0.4$\\
Arm 1 & $1.0 \pm 0.1$ & $1.1 \pm 0.1$ & $1.3 \pm 0.4$\\
Arm 2 & $1.1 \pm 0.1$ & $1.2 \pm 0.2$ & $1.4\pm 0.6$\\
Interarm 1 & $0.9 \pm 0.1$ & $1.1 \pm 0.1$ & $1.2 \pm 0.3$\\
Interarm 2 & $0.9 \pm 0.1$ & $1.1 \pm 0.1$ & $1.2 \pm 0.3$
\enddata
\end{deluxetable}

 In the following, we discuss colors based on S11 only.
 For other colors, see Appendix \ref{sec:othercolors}.

\subsubsection{S7 vs S11}\label{sec:S7/S11}
 \citet{Sak07} presented the AKARI S7/S11 map of the nearby spiral galaxy NGC 6946 and found that 
the ratio is larger than unity in arm regions and otherwise in interarm regions.
 They interpreted this trend as PAHs are ionized around massive star forming 
regions and the $7.7\mu$m PAH feature thus becomes brighter.
 Consistently, the PAH band ratio of 7.7$\mu$m/11.3$\mu$m is found to be larger in 
arm regions from spectral data of ISO \citep{Vogl05,Gal08}.

 From the color maps in Figure \ref{fig:color_wa}, 
we have found that S7/S11 traces the global arm structure of M51a, 
which is consistent with the trend found in NGC 6946.
 However, the difference in average between arm and interarm regions 
is only comparable to rms as listed in Table \ref{tb:colors}.
 Figure \ref{fig:chist} also shows that S7/S11 histograms for arm and interarm regions overlap
and their difference is small compared to their width.
 We attribute this to two factors: arm width and feathers.
 First, if arms were defined by eye based on MIR images, they could be narrower than 
the current definition from 
the smooth stellar model image.
 As is obvious from Figure \ref{fig:color_wa}, a fraction of arms shows S7/S11 $< 1$.
 Second, S7/S11 in interarm feathers is similar to values in arms.
 This S7/S11 distribution thus indicates that the underlying gravitational structure 
does not totally alter the PAH properties.

\begin{figure}
\plotone{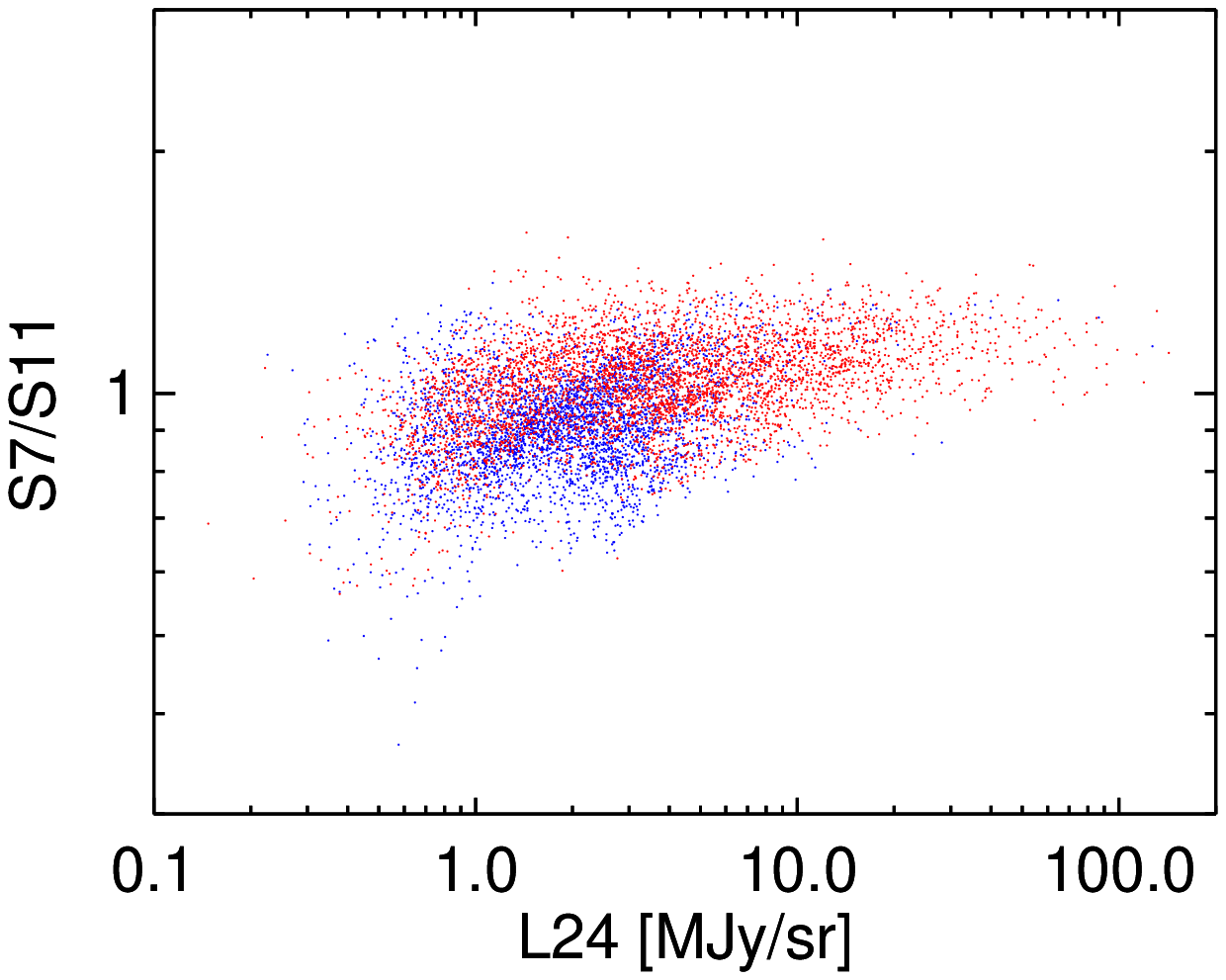}
\caption{S7/S11 ratio vs L24 surface brightness. Red is for arm while blue is for inter-arm regions.}
\label{fig:S7-S11_L24}
\end{figure}

 To investigate the cause of the S7/S11 variation, 
we plot the ratio vs L24 surface brightness in Figure \ref{fig:S7-S11_L24} 
with a hypothesis that star forming activities promote the PAH ionization.
 The correlation is however not so tight, and in fact in the arm regions, the ratio stays almost flat 
with respect to the L24 brightness.
 This plot indicates that the activeness of star formation 
is not the dominant factor deriving the S7/S11 variation.
 Supporting this result, from Spitzer/IRS spectra 
\citet{SmiJ07} found a nearly constant PAH band ratio $L(7.7)/L(11.3)$ 
across about two orders of magnitude in [\ion{Ne}{3}]/[\ion{Ne}{2}], 
an indicator of radiation hardness, for star forming nuclei.
 \citet{Gor08} also presented that PAH band ratios of \ion{H}{2} regions in the nearby galaxy M101 
do not depend strongly on the radiation hardness, while the equivalent width of each band feature 
decreases under very hard radiation.
 
 We have also found that in the inner part of Arm 1 
the ratio is larger on the downstream side, while 
this is not true for Arm 2, where most peaks appear around the middle point.
 This difference indicates that 
the variation of PAH ionization condition 
with respect to the arm structure  
(i.e.\ evolution across the spiral arms) 
is different between the two arms. 
 \citet{Egu09} found that spatial offsets between the arms of molecular gas and \ion{H}{2} regions 
are well explained by the spiral density wave hypothesis \citep[e.g.][]{LS64} for Arm 1 
but not for Arm 2.
 These differences in properties of the two arms
imply that the effect of the tidal interaction with M51b is more significant in Arm 2.
 We should however note here that 
the arm properties (and thus their differences) highly depend 
on how arm regions are defined.

 As already noted above, no clear radial trend of S7/S11 has been found for the disk 
(Figure \ref{fig:color_rad}).
 In the central region ($r\leq 30''$), the ratio 
steeply decreases toward the nucleus, 
which is consistent with Spitzer/IRS results such that 
$L(7.7)/L(11.3)$ is small in the M51a nucleus \citep{SmiJ07}.
 Since the decreasing trend of $L(7.7)/L(11.3)$ with [\ion{Ne}{3}]/[\ion{Ne}{2}] 
is found for active nuclei, contrary to flat distribution for star forming nuclei, 
smaller or ionized PAHs responsible for the 7$\mu$m feature are 
thought to be selectively destroyed in the nucleus of M51a.

\subsubsection{L15 \& L24 vs S11}
 Spatial distribution of the other colors, i.e.\ L15/S11 and L24/S11, is more patchy 
as shown in Figure \ref{fig:color_wa}.
 A high ratio is found in individual star forming regions, which mostly reside within the arms, 
while the broad S7/S11 distribution traces the global arm structure.

 As is clearly seen in Figure \ref{fig:color_rad} and Table \ref{tb:colors}, 
the difference in the average colors between the arm and interarm regions is even less 
distinct compared to their profile widths.
 Meanwhile, histograms in Figure \ref{fig:chist} show that larger color values 
are more frequently seen in arm regions, especially in the case of L24/S11.
 These results indicate that the very high ratios in a small number of pixels 
contribute to the large dispersion but not to the average.

 Steep rise toward the center reflects the harsh environments around 
the active nucleus where the interstellar dust is heated to a very high temperature.

\subsection{Color-color diagram}\label{sec:cc}
\begin{figure*}
\plotone{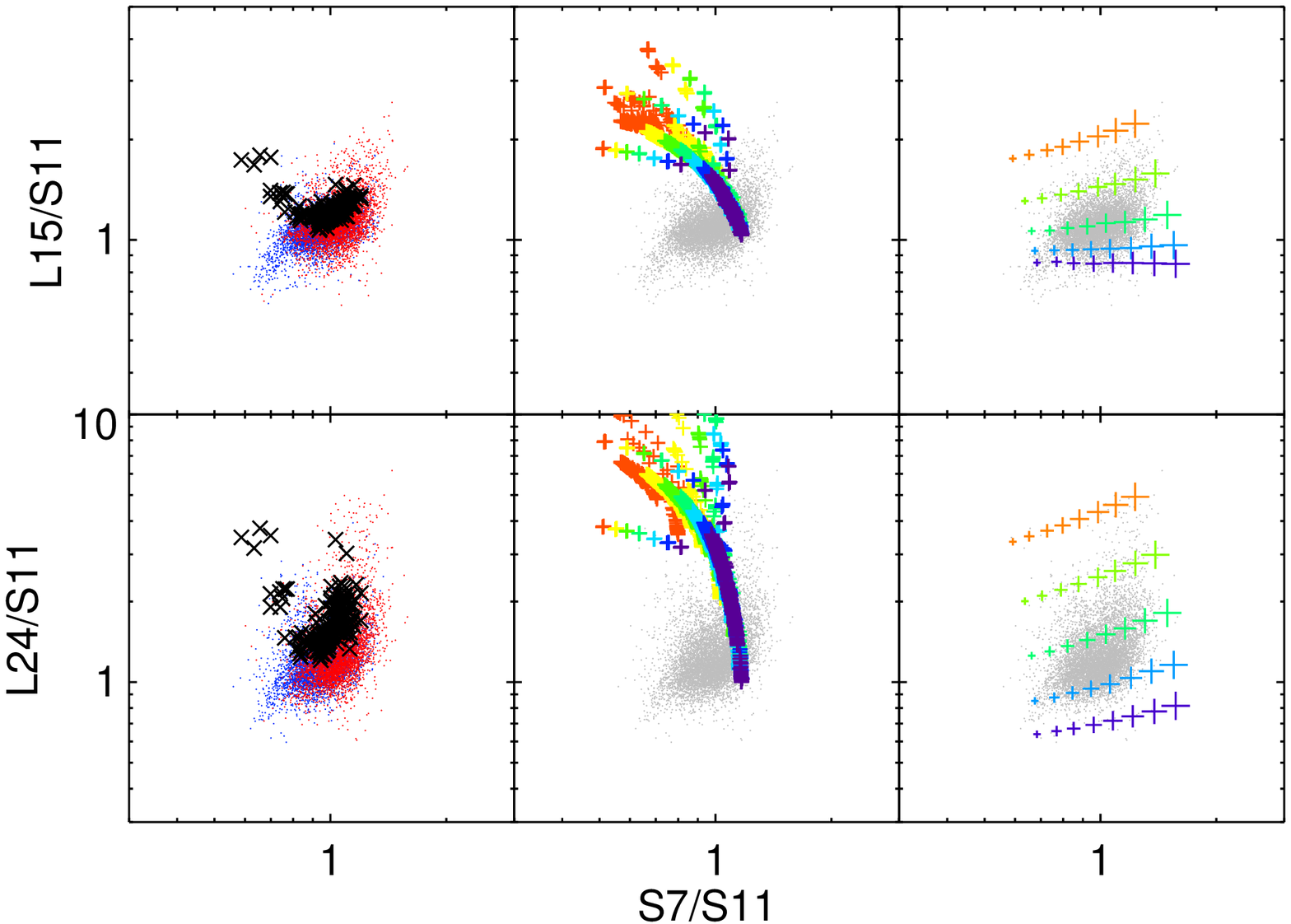}
\caption{Left: color-color diagrams of observational data color coded by the defined regions. 
(Blue and red dots represent interarm and arm regions, respectively, while black x is for the center.)
Middle: model predictions from \citet{DL07} indicated by crosses with color coded by 
the PAH fraction $q_{\rm PAH}$.
(Color changes from red to blue as $q_{\rm PAH}$ increases.)
Right: model predictions from this work indicated by crosses. 
(The symbol size and color represent $p_0$ and $p_1$ in Equation (\ref{eq:model}), respectively.)}
\label{fig:cc}
\end{figure*}

 In Figure \ref{fig:cc}, we present color-color diagrams.
 In the left two panels, data points are color coded by the defined regions -- 
red for arm, blue for interarm, and black for center.
 Since the two arm (interarm) regions do not differ significantly as presented above, 
we do not distinguish them hereafter.
 While most of the points for center overlap with arm points, 
four points reside in a distinct area, which corresponds to the active nucleus.

 In the middle two panels, model calculation from \citet[hereafter DL07]{DL07} 
is superimposed on data points in gray.
 The color represents the PAH mass fraction to the total dust $q_{\rm PAH}$, 
in the sense that it increases from red (0.47\%) to blue (4.58\%).
 The range of the other two variable parameters, 
the minimum threshold for radiation strength ($U_{\rm min}$) and 
the dust mass fraction exposed to $U_{\rm min} < U \leq U_{\rm max}$ ($\gamma$), 
was taken to be as wide as available: 
$\gamma=$0--1 and $U_{\rm min}=0.1$--$3\times 10^5$.
 Note that $U$ is in the unit of local MW radiation field.
 From these plots, it is clear that the observed color distribution is not fully covered 
by the combination of parameters employed in DL07.
 While \citet{DL01} presented a strong dependence of 
the band ratio $11.3\mu$m$/7.7\mu$m on the ionization state, 
DL07 adopted a typical PAH ionization fraction for the solar neighborhood ISM 
as a function of grain size.
 As a consequence, the profile of PAH band emission from DL07 models does not vary a lot 
according to the environments except in very strong radiation field (see their Figure 13), 
resulting in the very narrow range in predicted S7/S11 ratio.

 Given the importance of variation of the PAH ionization state as discussed above, 
we decide to construct a new simple model constituting neutral and ionized PAHs and 
larger dust grains with their relative abundance being independently variable.
 Since we only have MIR data where cold dust contribution is small, 
we use a modified blackbody spectrum with a single temperature 
$T=100$ K to represent the flux from larger dust grains 
and call it a warm dust (WD) component.
 The total emission from these three components are given by 
\begin{equation}
(1-p_0)\times I_\nu ({\rm PAH0}) + p_0 \times I_\nu ({\rm PAH1})
+ p_1 \times \kappa B_\nu (T=100 {\rm K}),
\label{eq:model}
\end{equation}
where $I_\nu ({\rm PAH0})$ and $I_\nu ({\rm PAH1})$ are the intensity 
for neutral and ionized PAHs from DL07 provided 
by the DustEM code \citep{Com11}, 
$\kappa$ is the dust emissivity function, 
and $B_\nu$ is the Planck function.
 Each of the three dust components is displayed in Figure \ref{fig:rsr}.

 We calculate the in-band flux of these dust components by 
convolving these spectra with the filter response curve of AKARI MIR bands, 
which are also presented in Figure \ref{fig:rsr}.
 By varying the two free parameters, $p_0$ and $p_1$, i.e., 
the fraction of ionized PAHs and the WD component, 
predicted colors from our model are plotted in the right panels of Figure \ref{fig:cc}.
 While the symbol size corresponds to $p_0$, 
the symbol color represents $p_1$, 
changing from blue to red as it increases.
 Overall, the WD fraction and the ionized PAH fraction are positively correlated, 
which indicates that when dust grains are warmer more PAHs are ionized. 
 This trend is consistent with \citet{Mori12}, 
who presented a positive correlation between 
the temperature of heating source and the ionization fraction of PAHs 
for selected positions in the Large Magellanic Cloud.

The S7/S11 distribution is now well explained by the variation of 
$p_0$ from 0\% to 70\%.
 The symbols in Figure \ref{fig:cc} are plotted every 10\%.
 The other parameter, $p_1$, is related to $1/q_{\rm PAH}$ 
in the sense that L15/S11 and L24/S11 increase as they increase.
 However, note that they are not exactly the same,
since $p_1$ is the ratio of the warm dust components to PAHs 
while $q_{\rm PAH}$ is the PAH fraction to total dust 
including colder components.
 By adopting the absorption opacity 
$\kappa = 2.92 \times 10^5 (\lambda/\mu {\rm m})^{-2}$ [cm$^2$/g] 
from \citet{LD01}, the dynamic range of $p_1$ corresponds to 
$M_{\rm WD}/M_{\rm PAH}=(0.83$--$13)\times 10^{-5}$.
 The symbols in Figure \ref{fig:cc} are plotted 
by changing this parameter by a factor of two.

\section{Discussion}\label{sec:dis}
 Based on the results presented above, we discuss the property of interstellar dust 
in the disk of M51a.

\subsection{PAH ionization}

 From the S7/S11 distribution map (Figure \ref{fig:color_wa}), 
we find S7/S11 generally traces the global spiral structure of M51a.
 Though its difference between the arm and interarm regions is only comparable 
to rms of their distribution, 
its variation cannot be reproduced by the dust model parameters provided by DL07 (Figure \ref{fig:cc}).
 We attribute this discrepancy to the variation of PAH ionization state, 
which is not fully implemented in their model.
 Our PAH+WD model suggests a significant variation of the ionized PAH fraction
(0--70\%) within the disk of M51a.

 The balance of ionization and recombination of PAHs 
is controlled by $G_0\sqrt{T_{\rm gas}}/n_e$,  
where $G_0$ is the radiation field strength, 
$n_e$ is the electron density, and $T_{\rm gas}$ is the gas temperature 
\citep[e.g.,][]{Tie08}.
 However, these parameters have been determined only in a few Galactic regions, 
so that their correlation with the relative strength of PAH features is 
investigated indirectly for extragalactic sources \citep{Gal08}.
Since all of these three parameters are supposed to become larger 
in active regions such as galactic nuclei and \ion{H}{2} regions, 
the PAH ionization is expected to proceed in arm regions, 
where star formation is more active than in interarm regions.

 Other candidates than the PAH ionization to change the PAH band ratio or S7/S11 color 
are the PAH size distribution and the shape of radiation field spectra, 
both of which are investigated by \citet{Gal08} and \citet{Dra11}.
 For the former, S7/S11 should become smaller when small size PAHs are less abundant.
 One indicator for the size distribution of PAHs is 
the ratio of $6.2\mu$m and $7.7\mu$m band features.
 \citet{Gal08} found this ratio relatively constant in star forming galaxies, 
which suggests that the size distribution does not significantly vary 
under the conditions of normal star formation.
 It is thus unlikely that the PAH size distribution 
is responsible for the difference in S7/S11 between the arm and interarm regions.
 For the latter, S7/S11 should become larger when the radiation is harder.
 Although the enhancement of the PAH 7.7$\mu$m/11.3$\mu$m ratio due to 
the radiation from young stars is consistent with large S7/S11 values in spiral arms, 
its effect compared to the normal radiation field is estimated to range 
from a few to a few tens of percent, which is not enough to explain 
the observed variations \citep{Gal08,Dra11}.

 As already mentioned in \S \ref{sec:S7/S11}, 
neither of the gravitational potential nor star formation rate 
can solely explain the S7/S11 variation.
 We therefore conclude that the observed variation in S7/S11 is mainly due to 
different ionization conditions of PAHs, though the physical conditions 
triggering their difference is unclear.

\subsection{S11 excess?}
 From the comparison with observational data and DL07 models 
(middle panels of Figure \ref{fig:cc}), we find that observational data points with 
L15/S11 $<1$ and L24/S11 $< 1$ do exist 
while the models only predict these ratios larger than 1.
 In the following, we consider the contribution to the S11 flux from 
components other than interstellar dust, 
which could result in this discrepancy.

 Since the contribution from stellar continuum has already been subtracted 
in \S \ref{sec:submodel}, another
candidate is gas emission lines. 
 \citet{Dale09a} obtained MIR spectra from Spitzer/IRS toward selected positions in M51 
(mostly star forming regions and nuclei) and measured the strength of these lines.
 Within the S11 spectral coverage, [\ion{Ne}{2}] at $\lambda=12.81 ~\mu$m is the 
strongest but its contribution to the S11 flux is estimated to be 3\% at most.
 In addition, L15/S11 $<1$ and L24/S11 $< 1$ are mostly found in 
interarm regions (blue points in the left panels of Figure \ref{fig:cc}), 
where emission lines are expected to be weaker.

 Therefore, we conclude that the discrepancy between observations and 
the DL07 model still remains unaccounted for.

\subsection{PAH fraction}
 As discussed above, we find an excess in S11 flux compared to DL07 models.
 This excess may indicate that in M51a the PAH fraction ($q_{\rm PAH}$) is larger than 4.58\%, 
which is the largest value employed in their model. 
 This is an important observational indication to their model, 
pointed out for the first time by this work since their model has mostly been applied to 
data from Spitzer, which does not have a photometric filter at 11 $\mu$m.

 Note again that the relative abundance of WD to PAHs employed in our model ($p_1$)
does not perfectly correspond to the PAH fraction discussed here.
 In order to discuss the fraction of each component including colder dust and its spatial variation, 
data at longer wavelengths (i.e.\ FIR and mm) with a high spatial resolution are needed 
and determining $q_{\rm PAH}$ is beyond the scope of this paper.

\subsection{Very small grains?}
 In the right panels of Figure \ref{fig:cc}, we plot predicted colors from the PAH+WD model 
over observed colors.
 While the two parameters are chosen to cover the observed distribution of 
L24/S11 vs S7/S11 (the bottom right panel), the L15/S11 distribution (the top right panel) is 
also covered with the same parameter range.
 This result indicates that one $T=100$ K component is enough 
to explain the AKARI MIR colors, and that hotter dust components 
such as very small grains (VSGs) stochastically heated to a very high temperature 
are not required.
 On the other hand, \citet{SmiJ07} 
employed multiple dust temperature components to fit 
the observed IR spectra of nearby star forming galaxies.
 This apparent inconsistency might be due to difference in the observing method 
(imaging and spectroscopy) and the wavelength coverage.
 However, we should note that our result does not fully reject 
multiple temperature components and the presence of VSGs.

\subsection{Extinction effects}
 We here estimate effects from dust extinction to observed band flux, 
which have been neglected so far.
 Foreground extinction is estimated from the Galactic dust distribution \citep{Schl98}.
 At the position of M51a, $A_V = 0.115$ mag.
 Extinction from dust within the galaxy is more difficult to derive. 
 \citet{Cal05} measured the amount of extinction toward \ion{H}{2} regions in M51a
with $13''$ aperture and the median value is about $A_V \simeq 2.6$ mag for 
$r\simeq 8''$--$80''$.
 While \citet{Pres07} found that the extinction in H$\alpha$
decreases by 0.9 magnitude over the optical radius of M51, 
\citet{Boi07} and \citet{Muno09} found similar gradients for
the extinction in the UV.
 These radial gradients indicate even smaller extinction in the outer disk.
 
 From models provided by \citet{Dra03}, $A(\lambda)/A_V\sim 0.01$--0.05 
for the four representative wavelengths of MIR bands.
 We thus estimate that the in-band flux is underestimated by $\sim$ 3--13\% 
due to extinction.
 We should note here that the extinction effect cannot explain the S11 excess, 
since it is largest in S11 due to the silicate absorption feature and the extinction correction 
would even more increase the S11 excess.
 Furthermore, the excess is mostly observed in the interarm regions where 
the extinction is plausibly less than other regions.

\section{Summary}\label{sec:sum}
 We present four MIR band images of the nearby pair of galaxies M51 
taken under two pointed observations with IRC on board AKARI.
 The field of view of IRC is about $10'$, covering almost the entire M51 system.
 Additional reduction procedures besides the standard pipeline are also summarized, 
which are fully described in Egusa et al.\ (2013, in prep.).
 The angular resolution is set to be $7.4''$ ($\sim 300$ pc) in FWHM after the PSF correction.

 A contribution from stellar continuum to each AKARI MIR band is 
estimated from the models by \citet{MenC12} and subtracted from the observed image 
to obtain a dust only image.
 Following \citet{Dum11}, we apply the wavelet analysis to the stellar model image 
and define arm and interarm regions in the disk of the main spiral galaxy M51a.

 The arm/interarm contrast is found to be slightly different between the four MIR bands. 
 The lowest contrast for S11 is interpreted that 
among the four AKARI MIR bands this band best represents the total dust distribution 
including colder components, while L24 with the highest contrast traces warmer dust heated 
from star forming activities.

 The flux ratio between the bands, i.e.\ color, has been measured over the disk of M51a. 
 As already found in another spiral galaxy NGC 6946 by \citet{Sak07}, 
S7/S11 well traces the global spiral arm pattern with higher value in the arms.
 However, its difference between arm and interarm regions is found to be 
only comparable to rms in their distribution.
 Furthermore, S7/S11 does not tightly correlate with L24 brightness, 
an indicator of star formation.
 These results indicate that the variation of S7/S11 does exist but none of 
the underlying stellar potential and activeness of star formation are the dominant controlling factor.

 Comparison with dust models by \citet{DL07} also suggests that 
the variation of PAH ionization state needs to be taken into account. 
 Employing a new model, which comprises ionized and neutral PAHs and $T=100$ K dust grains 
with their relative abundances being variable, we successfully cover the observed distribution 
in the color-color plots.
 The observed S7/S11 variation is explained by changing the ionized PAH fraction 
from 0\% to 70\%.
 The fraction of ionized PAHs and warm dust is positively correlated in the disk of M51a, 
indicating that when dust grains are warmer more PAHs are ionized.
 By comparing observed colors and the models by \citet{DL07}, 
we also find excess in the S11 surface brightness, which cannot be explained by 
stellar continuum, gas emission lines, and extinction effect.
 This excess indicates that the PAH fraction to total interstellar dust is likely higher than 
previous estimates in M51a.

 We emphasize that data at 11 $\mu$m are essential for the above discussions 
and thus to investigate PAH properties.

\acknowledgments
 The authors greatly appreciate help from Dr.\ Mentuch Cooper, 
who kindly provided the stellar model images.
 This work is based on observations with AKARI, 
a JAXA project with the participation of ESA.
 The authors are grateful to all the members of the AKARI project, 
and particularly to the IRC members for their helpful comments and 
fruitful discussion with them.
 This research has made use of the NASA/IPAC Extragalactic Database (NED) 
and NASA/IPAC Infrared Science Archive (IRSA), 
which are operated by the Jet Propulsion Laboratory, California Institute of Technology, 
under contract with the National Aeronautics and Space Administration.

{\it Facilities:} \facility{AKARI}

\appendix

\section{Comparison of colors with the literatures}\label{sec:othercolors}
 Here we present the distribution of colors not based on S11 and compare AKARI results with 
previous studies.
 As well as colors discussed in \S \ref{sec:colors}, the $5\sigma$ threshold has been employed
when calculating colors.

\begin{figure}
\includegraphics[scale=0.45,trim=60 10 60 30,clip]{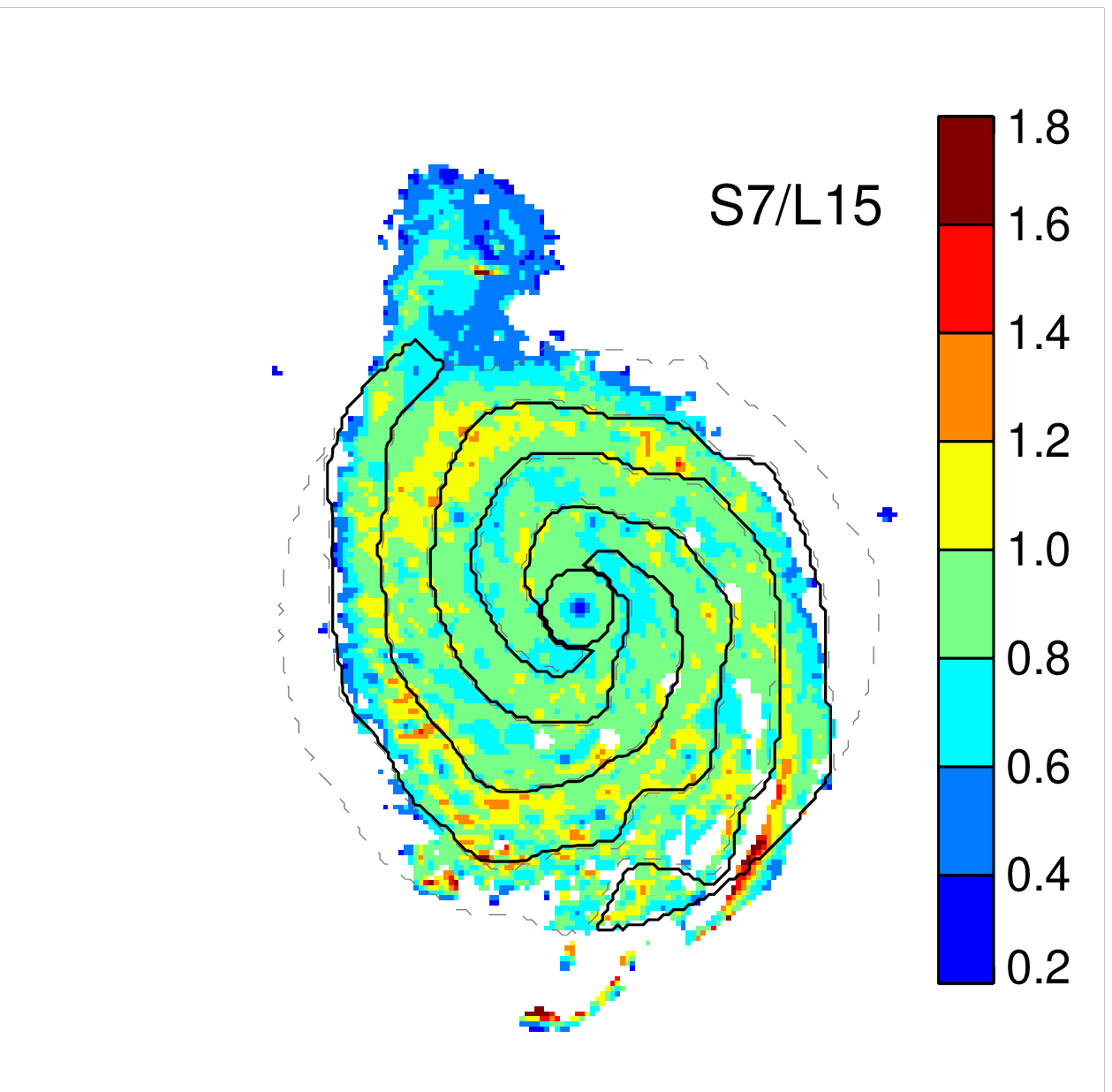}
\includegraphics[scale=0.45,trim=60 10 00 30,clip]{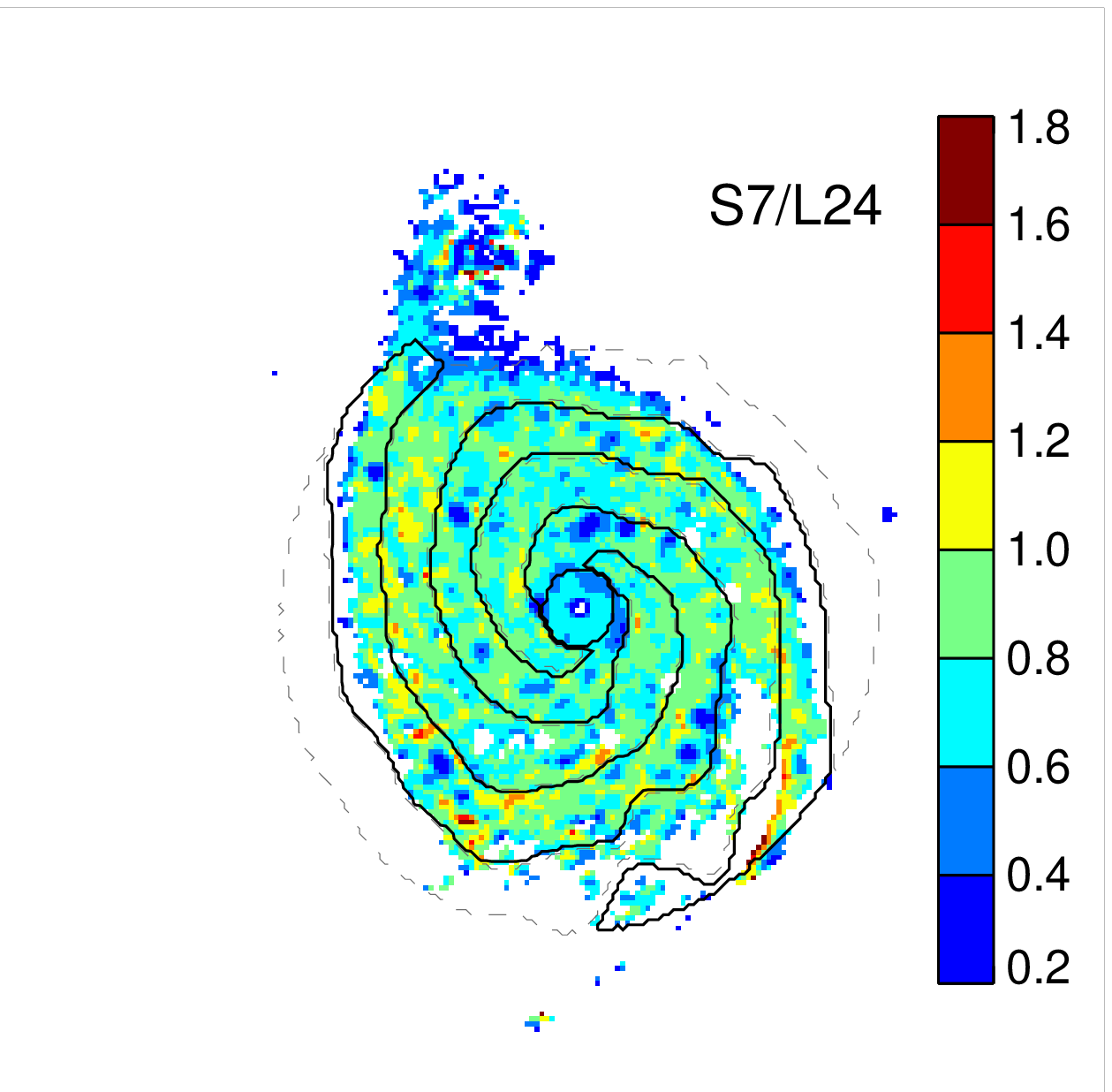}
\caption{Same as Figure \ref{fig:color_wa} but for S7/L15 and S7/L24.}
\label{fig:ocolor_wa}
\end{figure}

\subsection{S7 vs L15}
 \citet{Sauv96} analyzed LW2 ($7\mu$m) and LW3 ($15\mu$m) images of M51 from ISO
and found that LW2/LW3 becomes small in arm regions.
 They interpreted this result as a large dependence of LW3 flux on temperature; 
away from an arm, temperature becomes low resulting in high LW2/LW3 ratio 
in interarm regions.

 In contrast to the S7/S11 map, AKARI S7/L15 map shown in Figure \ref{fig:ocolor_wa}
is complicated.
 At positions \citet{Sauv96} derived the LW2/LW3 ratio, 
the S7/L15 ratio is slightly lower in the arm compared to the adjacent interarm region.
 The trend however does not seem to be universal.
 On average, we do not find any systematic difference in S7/L15 for arm and interarm regions.
 
 \citet{Rou01} measured LW3/LW2 for selected bright objects in M51,
mostly star forming regions and the companion, and showed that 
the ratio is almost constant around 1.0--1.5 with no significant radial trend.
 A roughly constant LW2/LW3 color was also found in 
the disk of NGC 6946 \citep{Helou96}.
 Consistently, we confirm that S7/L15 values for both arm and interarm regions 
do not show any radial dependence, while it steeply decreases toward the nucleus in the center region.

\subsection{S7 vs L24}
 Bright PAH band features around 7 $\mu$m and 
the continuum emission around 24 $\mu$m are 
often regarded as a star formation indicator.
 However, their spatial distribution around \ion{H}{2} regions 
has been found to differ.
 By subtracting the stellar contribution from Spitzer IRAC4 (8 $\mu$m) images 
and comparing them with MIPS24 (24 $\mu$m) images, 
many authors have reported that 
the stellar subtracted IRAC4, which is dominated by the PAH features, 
is bright around the border of \ion{H}{2} regions, 
while MIPS24 peaks at their center \citep[e.g.][]{Helou04,Ben06}.
 The wavelet analysis by \citet{Dum11} also supports this finding in the sense that 
the amplitude shows a peak around $10''$ scale for MIPS24 while it is flat 
around that scale for stellar subtracted IRAC4. 
 From these results, PAH grains are regarded to be destroyed in the center of \ion{H}{2} regions, 
where the stellar radiation is hard and strong.

\begin{figure}[!t]
\plotone{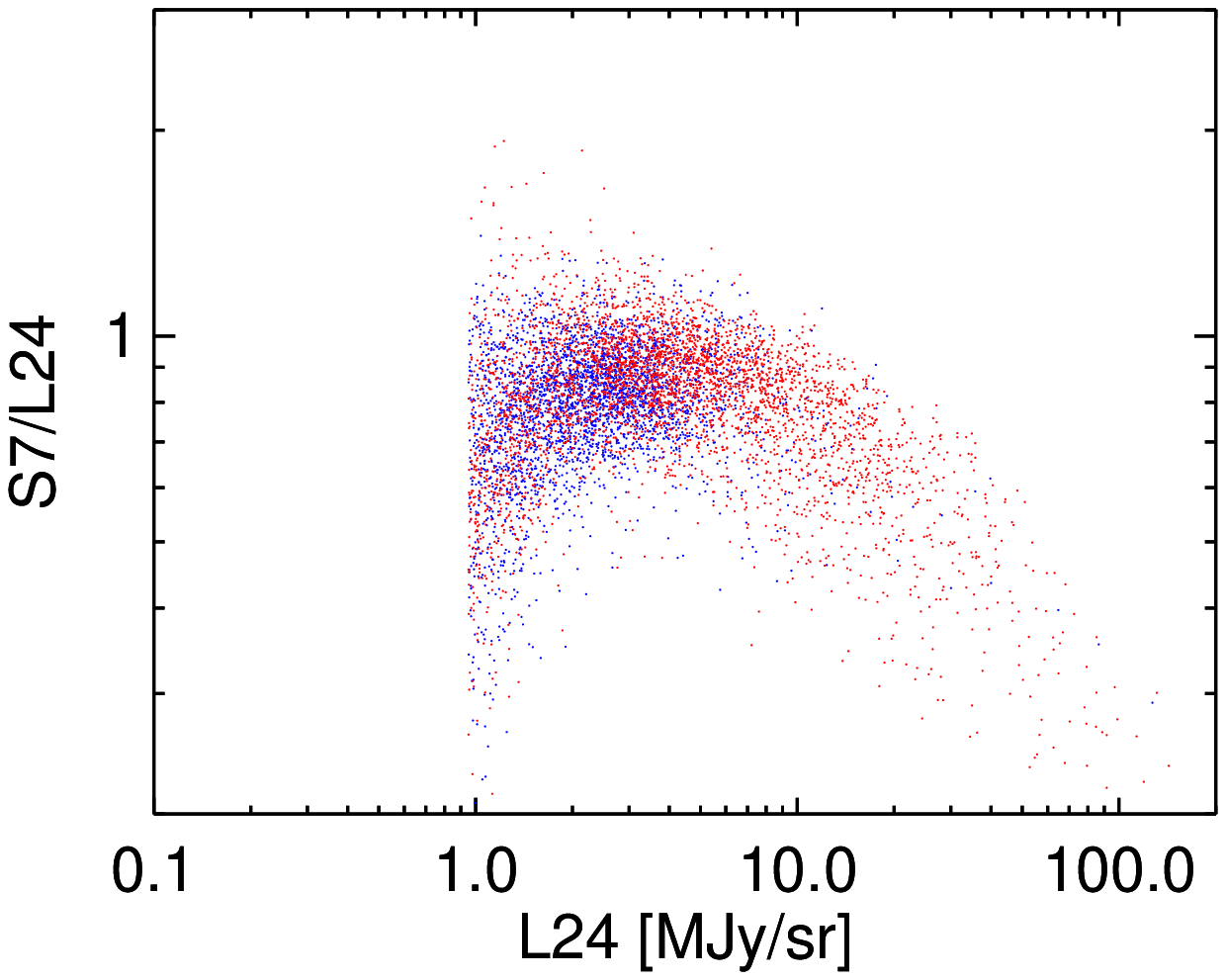}
\caption{S7/L24 against L24 surface brightness. 
Red and blue points represent arm and interarm regions, respectively.
The cutoff at the faint end corresponds to the $5\sigma$ threshold 
applied when calculating S7/L24.}
\label{fig:S7/L24_L24}
\end{figure}

 We also find a consistent trend in our AKARI S7/L24 map (Figure \ref{fig:ocolor_wa}), 
in which the ratio is small in bright \ion{H}{2} regions in spiral arms.
 In order to investigate its relationship to star formation rate, S7/L24 is plotted 
against L24 surface brightness in Figure \ref{fig:S7/L24_L24}.
 At the bright end, i.e.\ around \ion{H}{2} regions, the ratio decreases with increasing L24, 
which is consistent with the above picture.
 At the faint end, on the other hand, it stays mostly flat or even increases with L24, 
though the scatter is large.
 We deduce that this bimodal relationship reflects multiple origins of the L24 flux; 
while young massive stars are the dominant source for bright objects, old and less massive 
stars contribute significantly where massive star formation is absent.
 The turning point of these two trends is L24 $\sim$ 3--5 MJy/sr, 
which is around the mode of L24 surface brightness distribution for both arm and interarm regions 
(Figure \ref{fig:fcomp}).

 We again do not find any systematic difference
in average S7/L24 between the arm and interarm regions.
 Radial dependence is also absent except the central region, where it steeply decreases 
toward the nucleus.

\bibliography{papers}

\end{document}